\documentclass{aa}
\usepackage{multirow}
\usepackage{rotating}
\usepackage{bmpsize}
\usepackage{graphicx}
\usepackage{epsfig}
\usepackage[]{natbib}
\bibpunct{(}{)}{;}{a}{}{,} 
\usepackage{txfonts}
\usepackage{psfrag}
\usepackage{units}
\def\Msun{$M_{\sun}$}

\def\nh{{n_{\rm H}}}
\def\nh2{{n(\rm H_2)}}
\def\h2{${\rm H_2}$}

\def\3cm{\rm {cm^{-3}}}
\def\2cm{\rm {cm^{-2}}}
\def\s-1{\rm {s^{-1}}}
\def\etal {et al.}
\def\mum {\hbox{$\mu$m}}
\def\kms {\hbox{${\rm km\,s}^{-1}$}}
\def\Kkms {\hbox{${\rm K}\,{\rm km}\,{\rm s}^{-1}$}}
\def\ndv{\hbox{${\rm cm}^{-2}\,{\rm km}^{-1}\,{\rm s} $}}

\def\co{\hbox{\rm {CO}}}
\def\twco{\hbox{{\rm $^{12}$CO}}}
\def\thco{\hbox{{\rm $^{13}$CO}}}
\def\co17{\hbox{{\rm {C}${}^{17}$O}}}
\def\ceio{\hbox{{\rm {C}${}^{18}$O}}}
\def\oi{\hbox{{\rm [O {\scriptsize I}]}}}
\def\ci{\hbox{{\rm [C {\scriptsize I}]}}}
\def\cii{\hbox{{\rm [C {\scriptsize II}]}}}
\def\nii{\hbox{{\rm [N {\scriptsize II}]}}}
\def\h1{\hbox{{\rm H {\scriptsize I}}}}
\def\hii{\hbox{{\rm H {\scriptsize II}}}}
\def\hh{\hbox{H$_2$}}
\def\pb{P\'{e}rez-Beaupuits}

\begin{document}

\title{Detection of a large fraction of atomic gas not associated with star-forming material in M17~SW}
\author{J.P. P\'{e}rez-Beaupuits\inst{1} \and
        J.~Stutzki\inst{2} \and
        V.~Ossenkopf\inst{2} \and      
        M.~Spaans\inst{3} \and  
        R.~G\"usten\inst{1} \and
        H.~Wiesemeyer\inst{1} 
}
\offprints{J.P. P\'erez-Beaupuits}
\institute{
 Max-Planck-Institut f\"ur Radioastronomie, Auf dem H\"ugel 69, 53121 Bonn, Germany -
 \email{jp@mpifr.de}
 \and
 I. Physikalisches Institut der Universit\"at zu K\"oln, Z\"ulpicher Stra\ss e 77, 50937 K\"oln, Germany 
 \and 
 Kapteyn Astronomical Institute, Rijksuniversiteit Groningen, 9747 AV Groningen, The Netherlands
}
\date{Received  / Accepted  }
\titlerunning{Atomic Gas not associated in M17~SW}

%
\abstract
  {The \cii~158~\mum\ line is one of the dominant coolants of the ISM, and an important probe with which to study the 
  star formation process. Recent Herschel/HIFI and SOFIA/GREAT observations showed that assuming the total 
  velocity-integrated intensity of this line is directly associated with the star-forming material is 
  inadequate. 
  }
  {We  probe the column densities and masses traced by the ionized and neutral atomic carbon with spectrally 
  resolved maps, and compare them to the diffuse and dense molecular gas traced by \ci\ and low-$J$ CO lines toward 
  the star-forming region M17~SW.
  }
  {We mapped a 4.1 pc $\times$ 4.7 pc region in the \ci~609~\mum\ line using the APEX telescope, as well as the 
  CO isotopologues with the IRAM 30m telescope. 
  Because of the velocity-resolved spectra, we analyze the data based on  velocity channel maps that are 1~\kms\ wide.
  We correlate their spatial distribution with that of the \cii\ map obtained with 
  SOFIA/GREAT. Optically thin approximations were used to estimate the column densities of \ci\ and \cii\ in 
  each velocity channel. 
  }
  {
  The distribution of the emission from the isotopologues \thco, \co17, and \ceio\ resembles more 
  closely that of the \ci\ emission than that of the \twco\ emission. The spatial distribution of the \ci\ and all CO 
  isotopologues emission was found to be associated with that of \cii\ in about 20\%--80\% of the mapped region, 
  with the high correlation found in the central (15--23~\kms ) velocity channels.
  }
  {The excitation temperature of \ci\ ranges between 40~K and 100~K in the inner molecular
  region of M17~SW. Excitation temperatures up to 200~K are found along the ridge.
  Column densities in 1~\kms\ channels between $\sim$10$^{15}~\2cm$ and $\sim$10$^{17}~\2cm$ were found 
  for \ci. Just $\sim$20~\% of the velocity range ($\sim$40~\kms) that the \cii\ line spans is associated with 
  the star-forming material traced by \ci\ and CO. 
  The total (integrated over the 0--40~\kms\ velocity range) gas mass estimated from 
  the \cii\ emission gives a lower limit of $\sim$4.4$\times$10$^3$~\Msun. A very large fraction of at 
  least 64\% of this mass is not associated with the star-forming material in M17~SW. We also 
  found that about 36\%, 17\%, and 47\% of the \cii\ emission is associated with the \hii, 
  \h1, and \hh\ regimes, respectively. Comparisons with the H41$\alpha$ line shows an ionization 
  region mixed with the neutral and part of the molecular gas, in agreement with the clumped structure and 
  dynamical processes at play in M17~SW. These results are also relevant to extra-galactic studies in which 
  \cii\ is often used as a tracer of star-forming material.
  }


\maketitle

\section{Introduction}

In order to advance our understanding of the ambient conditions of star formation, observations of large areas 
of known massive Galactic star-forming regions have been done over a wide range of wavelengths.
Observations of low- and mid-$J$ transitions of \twco\ towards several massive star-forming regions
have shown that  warm and dense gas is usually confined to narrow ($<1$ pc) zones close to the ionization 
front \citep[e.g.,][]{harris87, graf93, yamamoto01, kramer04, kramer08, pb10}. 
Although slow shocks and cloud-cloud collisions can be an important source of heating in high velocity wing 
objects like Orion, W51, and W49 \citep{jaffe87}, narrow mid-$J$ \twco\ lines, as well as the parameters needed 
to explain the CO observations, favor photoelectric heating of the warm gas located beyond 
the \hii\ region driven by the UV radiation field emerging from an ionizing source, the so-called photon-dominated region (PDR).

A vast amount of data also shows that molecular clouds are clumpy over a wide range of scales, from giant 
molecular clouds containing smaller subclouds, to small dense cores within the subclouds. The clumpiness of 
molecular clouds is relevant for the understanding of fragmentation processes that can lead to compact 
condensations that may collapse to form stars \citep[e.g.,][and references therein]{carr87, banerjee04, klessen05, hocuk10, clark11, federrath12}.
Several efforts have been made over the past years to identify clumps, their size, line width, and mass, among 
other parameters.
The complex line profiles observed in optically thin lines (e.g., \ci, CS, \ceio, \thco) and their velocity-channel maps, are indicative of the clumpy structure of molecular clouds and allow  a robust estimation of 
their clump mass spectra \citep[e.g.,][]{carr87, loren89, stutzki90, hobson92, kramer98, kramer04, pb10}.

Inhomogeneous and clumpy clouds, as well as a partial face-on illumination, in star-forming regions like M17~SW, 
NGC3603, S140, the Orion Molecular Cloud, and the NGC~7023 Nebula, are thought to produce extended emission of the atomic lines \ci\ and \cii, and suppress the stratification in \cii, \ci,\ and CO expected from standard 1-D steady-state PDR models, which is not observed in several sources
\citep[e.g.,][]{keene85, genzel88, stutzki88, spaans97, gerin98, yamamoto01, schneider02, roellig11, mookerjea03, pb10}.

Massive star-forming regions like the Omega Nebula M17, with an edge-on view (particularly in its 
southwest region), are ideal sources to study the clumpy structure of molecular clouds, as well as the chemical 
and thermodynamic effects of the nearby ionizing sources.
The southwest region of M17 (M17~SW) concentrates molecular material in a clumpy structure.
Models based on far-IR and submillimeter observations \citep{stutzki88, meixner92} suggest that 
the distribution and intensity of the emissions observed in the M17~SW complex, can be explained with high 
density ($n(\rm H_2)\sim5\times10^5~\3cm$) clumps embedded in an interclump medium ($n(\rm 
H_2)\sim3\times10^3~\3cm$) and surrounded by a diffuse halo ($n(\rm H_2)\sim300~\3cm$).

The central cluster of more than 100 stars that illuminates M17~SW is NGC~6618 \citep[e.g.,][]{lada91, hanson97}. 
The two components of the massive binary CEN1 \citep{kleinmann73, chini80} 
are part of the central cluster NGC~6618 and are separated by $\sim1\farcs 8$. This source, originally classified 
as a double O or early B system by Kleinmann (1973), is actually composed of two O4 visual binary stars, named 
CEN 1a (NE component) and CEN 1b (SW component), and it appears to be the dominant source of photo-ionization in 
the whole M17 region \citep{hoffmeister08}.

Recent SOFIA/GREAT observations of the velocity-resolved \cii\ spectra showed that a large fraction ($>$60\%) of 
the \cii\ emission, observed at the lower ($<$10~\kms) and higher ($>$24~\kms) velocity channels, is not 
associated with the star-forming material (denser and colder gas) traced by species like CO and 
\ci, which has an average line width of 5 to 10~\kms\ centered at $V_{LSR}=20$~\kms\ \citep{pb12, pb13}.
Only the central narrow (1~\kms) channel maps of the velocity-resolved \cii\ spectra show a spatial association 
with other gas tracers (e.g., \ci\ and \twco). The broader velocity range covered by the \cii\ 
line with respect to the \ci\ and \twco\ has to be associated with additional material, either lower density 
clumps or more diffuse, possibly ablated material, resulting in additional layers of ionized carbon 
gas within the telescope beam. 
The \cii\ emission have been found to extend at least $\frac{1}{4}^{\circ}$ in the sky \citep{russell81}, 
and $\sim$15 pc into the M17~SW molecular region \citep{stutzki88}. The spatial distribution of the \cii\ 
emission (and abundance) in the southern region of M17~SW does not follow theoretical predictions of stratified 
or clumpy PDR models \citep{pb12}.

In earlier works, high resolution maps of high- and mid-$J$ CO lines, the $^3P_2\rightarrow~^3P_1$ 
fine-structure transition of \ci, and the \cii~158~\mum\ emission, have been reported
\citep{pb10, pb12}.
In this study we present a new high resolution map of the $^3P_1\rightarrow~^3P_0$ fine-structure transition of 
\ci, as well as maps of the $J=1\rightarrow0$ and $J=2\rightarrow1$ transitions of \twco\ and its isotopologues.
In contrast to \cii~158~\mum\ (and \oi\ 63~\mum, not included in the present data set),
PDR models predict that the intensity of the \ci\ fine 
structure lines do not have a strong dependence on UV energy density \citep[e.g.,][]{hollenbach99}. 
Therefore, in a clumpy cloud irradiated by UV photons, the intensity of the \ci\ emission is expected to be 
proportional to the number of photodissociation surfaces of clumps along the line of sight 
\citep[e.g.,][]{ spaans96, howe00, kramer04}. Since several velocity components along the line of sight can be 
found in a clumpy medium, we present our analysis and discussions of the new results based on velocity channel 
maps, showing the temperatures of the lines integrated over a narrow 1~\kms\ channel width. From them we estimate 
the excitation temperature and column density of \ci, as well as the column density of \cii\ and the gas mass not 
associated with the star-forming material traced by \ci\ and the CO isotopologues.

The organization of this article is as follows. In Sect. 2 we describe the observations. The maps of the observed 
lines are presented in Sect. 3. The excitation temperature and column densities, as well as mass estimates, are 
presented in Sect. 4. In Sect. 5 we estimate the \cii\ emission not associated with other gas tracers. The 
conclusions and final remarks are presented in Sect. 6.

\section{Observations}\label{sec:observations}

\subsection{The APEX data}

We  used the higher frequency band of the dual channel receiver FLASH \citep[][hereafter FLASH-460]{heyminck06} 
on the Atacama Pathfinder EXperiment (APEX\footnote{This publication is based on data acquired with the Atacama 
Pathfinder Experiment (APEX). APEX is a collaboration between the Max-Planck-Institut f\"ur Radioastronomie, the 
European Southern Observatory, and the Onsala Space Observatory}; \citealt{gusten06}) during October 2009 to map 
the ${}^3{P_1}\to{}^3{P_0}$ 609~\mum\ (hereafter $1\to0$) fine-structure transition of \ci\ at 492.161 
GHz.
The observed region covers about $6'.2\times7'.2$ (4.1 pc $\times$ 4.7 pc) compared to the $5'.3\times4'.7$ (3.4 
pc $\times$ 3.0 pc) area previously mapped for \ci~${}^3{P_2}\to{}^3{P_1}$ 370~\mum\ (hereafter $J = 
2\to1$) with CHAMP$^+$ \citep{pb10}.
The \ci~$J = 1\to0$ was observed in on-the-fly (OTF) slews in RA ($\sim360$ arcsec long). Because the 
beam size of APEX at 492~GHz is about $12\farcs 7$, the subsequent scans in Declination and RA were spaced $6''$ 
apart.

The total power mode was used for the observations, nodding the antenna prior to each OTF and raster
slew to an off-source position ($180''$,$0''$), east of the star SAO~161357. This is used as the reference 
position ($\Delta \alpha=0$, $\Delta \delta=0$) in the maps and throughout the paper, with 
RA(J2000)=18:20:27.64 and Dec(J2000)=-16:12:00.90. The OFF position at $180''$ was determined to be clean, even in the \ci~$J=2\to1$ and mid-$J$ \twco\ lines.
The reference for continuum pointing was Sgr B2(N) and the pointing accuracy was better than $3''$ for all the 
maps.
The data were processed during the observations with the APEX real-time calibration software 
\citep{muders06}, assuming equal gains for the signal and image sidebands.

A fast Fourier transform spectrometer (FFTS), providing 1.5 GHz bandwidth and 2048 channels \citep{klein12}, was 
used for the \ci~$J = 1\rightarrow0$ map. 
The on-source integration time per dump was 1~second for the OTF map of \ci~$J = 1\rightarrow0$,
and the average DSB system noise temperature of the FLASH-460 was about 810~K. 

Observations toward Jupiter were performed during October 2009 to estimate the beam coupling efficiency 
($\eta_c\approx0.59$) of the FLASH-460, assuming a brightness temperature of 158~K for 
this planet at 492~GHz, as interpolated from data reported in \citet{griffin86}.

\begin{figure}[!hpt]

  \hfill\includegraphics[angle=0,width=8cm]{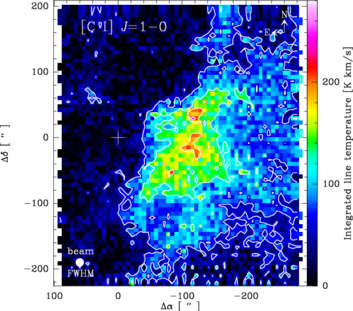}\hspace*{\fill}\\

  \hfill\includegraphics[angle=0,width=8cm]{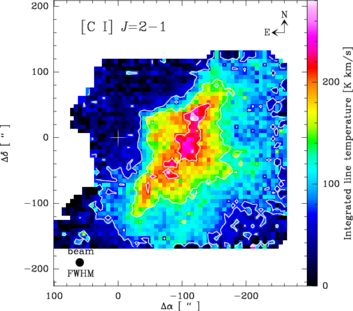}\hspace*{\fill}\\

  \caption{\footnotesize{\textit{Top} - Color map of the velocity integrated (in the range 0--40~\kms) 
  intensity of \ci\ $J = 1\rightarrow0$ in M17~SW. The peak emission is 240~\Kkms. The contour levels are 25\%, 
  50\%, 75\%, and 90\% of the peak emission. \textit{Bottom} - Color map of the integrated intensity of 
  \ci\ $J = 2\rightarrow1$ (from P\'erez-Beaupuits \etal\ 2010) convolved to the beam size 
  ($\sim12\farcs 7$) of the \ci\ $J = 1\rightarrow0$ line, with a peak emission of 280 \Kkms. The 
  contour levels are as described above. The reference position ($\Delta \alpha=0$, $\Delta 
  \delta=0$) is as in Fig.~\ref{fig:IRAM30m-maps1}.
}}

  \label{fig:APEX-maps}
\end{figure}

\subsection{The IRAM 30m data}

We used four frequency setups of the broadband EMIR receivers \citep{carter12} at IRAM 30m to map a similar 
area to the one mapped in \ci\ with the APEX telescope. 
The 32~GHz signal bandwidth provided by the IF channels were used for the receiver bands E090 (3mm) and E230 
(1.3mm), covering each sideband with 8~GHz bandwidth in single polarization. These setups allowed us to fully 
map all the CO isotopologues (in addition to many other molecules) in their $J=1\to0$ and $J=2\to1$ transitions. 
The beamwidths (FWHM) for the $J=1\to0$ transitions of \twco, \thco, \ceio, and \co17\ are 
$22\arcsec.6$, $23\arcsec.7$, $23\arcsec.8$, and $23\arcsec.2$, respectively. We also detected, by 
serendipity, the hydrogen recombination lines H39$\alpha$, H40$\alpha$, and H41$\alpha$ (28$\farcs$3) in the 3mm 
band.

The total region mapped of about 360$''\times$300$''$ was covered with two long OTF maps of 360$''\times$160$''$ 
(with an overlap of 20$''$ between them) and slews in RA ($\sim360$ arcsec long) with steps of 4$''$ in Dec. 
The off-source reference position was observed for 10~seconds every two OTF subscans (rows). The on-source 
integration time per dump was 0.5~second.

In order to ensure atmospheric stability, we used the same nearby off-source reference position 
(345$''$,-230$''$) as for the \cii\ map \citep{pb12} obtained 
with the German REceiver for Astronomy at Terahertz frequencies (GREAT\footnote{GREAT is a development by the 
MPI f\"ur Radioastronomie and the KOSMA/ Universit\"at zu K\"oln, in cooperation with the MPI f\"ur 
Sonnensystemforschung and the DLR Institut f\"ur Planetenforschung.}, \citealt{heyminck12}) on board  the 
Stratospheric Observatory For Infrared Astronomy (SOFIA). 
From previous APEX observations (not reported here) of the \twco~$J=3\to2$ and $J=4\to3$, we know that the 
reference position (345$''$,-230$''$) is not free of CO emission. Hence, for all the EMIR frequency setups, we 
first did a deep observation at the position (345$''$,-230$''$) against the reference that is even farther away  
at (3600$''$, -1800$''$) which, in turn, was verified to be CO emission-free against the offset position at 
(4600$''$,-2800$''$). Then we added the flux from the reference position (with about two orders of magnitude 
lower rms than the OTF spectra) back into the spectra from the OTF maps.

Beam coupling efficiencies for each individual line were obtained from interpolation of the values given in the 
online table of the IRAM 30m efficiencies\footnote{http://www.iram.es/IRAMES/mainWiki/Iram30mEfficiencies}. 
With these beam coupling efficiencies, and a forward efficiency $\eta_f$ (0.95 for EMIR090, 0.94 
for EMIR230, and 0.93 for EMIR150), we converted all data to the main beam brightness temperature scale, 
$T_{B}=\eta_{f}\times T_{A}^{*}/\eta_{c}$.
The reduction of these calibrated data, as well as the maps shown throughout the paper, were done using the 
GILDAS\footnote{http://www.iram.fr/IRAMFR/GILDAS} package CLASS90.

\begin{figure*}[!ht]

  \hfill\includegraphics[angle=0,width=7.0cm]{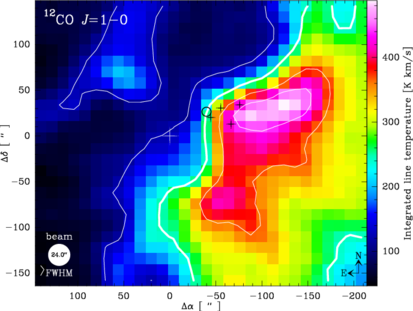}%
  \hfill\includegraphics[,angle=0,width=7.0cm]{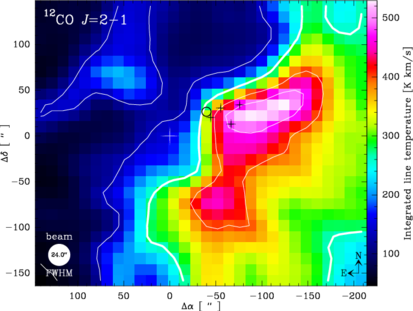}\hspace*{\fill}\\

  \hfill\includegraphics[angle=0,width=7.0cm]{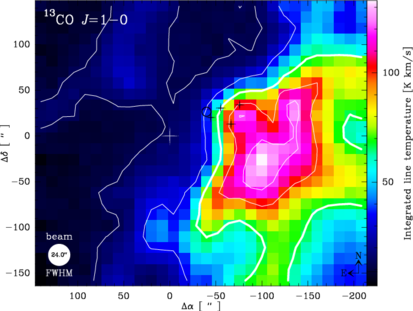}%
  \hfill\includegraphics[,angle=0,width=7.0cm]{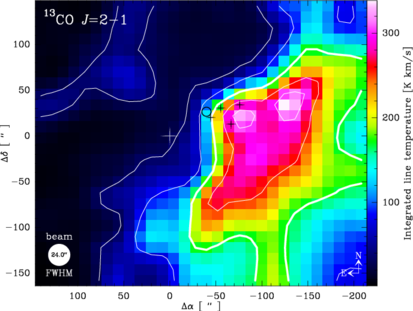}\hspace*{\fill}\\

  \caption{\footnotesize{Color maps of the velocity integrated intensity (in the range 0--40~\kms)  of the 
  $J = 1\rightarrow0$ and $J = 2\rightarrow1$ transitions of \twco\ (\textit{Top panels}) and \thco\ 
  (\textit{Bottom panels}) in M17 SW. 
  The contour levels are 25\%, 50\% (thick contour), 
  75\%, and 90\% of the peak emission. All maps have been convolved to the larger beam 
  (24$''$) of the \ceio\ $J = 1\rightarrow0$ line. The reference position ($\Delta \alpha=0$, $\Delta 
  \delta=0$), marked with a cross, corresponds to the SAO star 161357 at RA(J2000)=18:20:27.65 and 
  Dec(J2000)=-16:12:00.91. The ultracompact \hii\ region M17-UC1 and four H$_2$O masers \citep{johnson98} are 
  marked by the black circle and plus symbols, respectively.
}}

  \label{fig:IRAM30m-maps1}
\end{figure*}

\begin{figure*}[!ht]

  \hfill\includegraphics[angle=0,width=7.0cm]{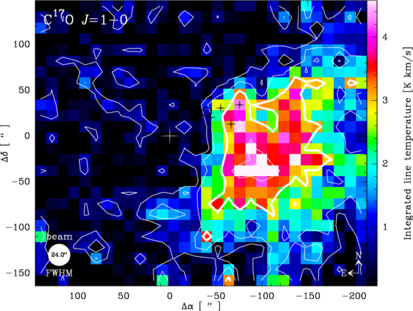}%
  \hfill\includegraphics[,angle=0,width=7.0cm]{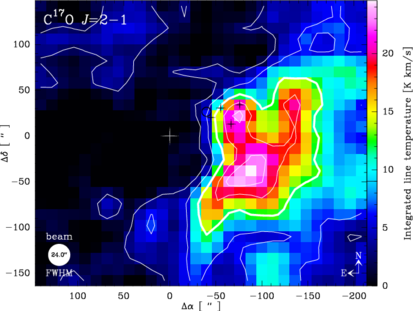}\hspace*{\fill}\\

  \hfill\includegraphics[angle=0,width=7.0cm]{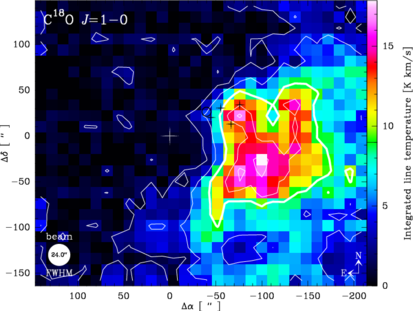}%
  \hfill\includegraphics[,angle=0,width=7.0cm]{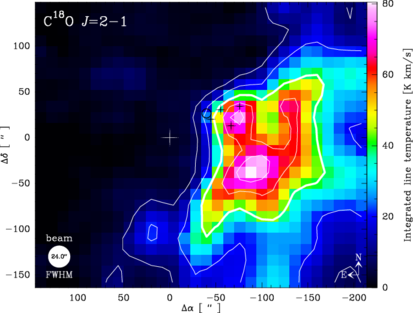}\hspace*{\fill}\\

  \caption{\footnotesize{Color maps of the velocity integrated intensity (in the range 0--40~\kms)  of the $J = 
  1\rightarrow0$ and $J = 2\rightarrow1$ transitions of \co17\ (\textit{Top panels}) and \ceio\ (\textit{Bottom 
  panels}) in M17 SW. Only the \co17~$J=1\to0$ line was integrated between 14 and 28 \kms\ because it is 
  fainter 
  and noisier than the other lines. The contour levels are 25\%, 50\% 
  (thick contour), 75\%, and 90\% of the peak emission. All maps have been convolved to the larger beam 
  (24$''$) of the \ceio\ $J = 1\rightarrow0$ line. The reference position ($\Delta \alpha=0$, 
  $\Delta \delta=0$) and symbols are as in Fig.~\ref{fig:IRAM30m-maps1}.
}}

  \label{fig:IRAM30m-maps2}
\end{figure*}

\subsection{SOFIA \cii\ observations}

The \cii~158~\mum\ map was already reported in \citet{pb12}, where a detailed explanation of the calibration and 
the OTF maps was given. Here we use the  same data as in the previous paper, convolved with 
a larger beam, and re-sampled the spectra to a 1~\kms\ channel width, as explained in 
Sect.~\ref{sec:CII-emission}. 

The SOFIA data is publicly available in the section ``Data Archive \& Retrieval'' of the SOFIA Data Cycle System\footnote{https://dcs.sofia.usra.edu/}. All the data presented in this work will be available as FITS files in the Strasbourg astronomical Data Center (CDS\footnote{http://cdsweb.u-strasbg.fr/}).

\section{Results}

\subsection{The \ci\ integrated intensity maps}\label{sec:CI-results}

Figure~\ref{fig:APEX-maps} shows the maps of the intensity, integrated between 0~\kms\ and 
40~\kms, of \ci~$J = 1\rightarrow0$ (\textit{top}) and $J = 2\rightarrow1$ (\textit{bottom}). 
Because the \ci~$J = 2\rightarrow1$ map was convolved to the larger beam size ($12\farcs 7$) of the 
$J = 1\rightarrow0$ transition, its peak integrated intensity is $\sim20$~\Kkms\ lower than the 
peak value previously reported in \citet{pb10}.
The peak integrated intensities of the maps shown here are 240~\Kkms\ and 280~\Kkms\ for the 
\ci~$J = 1\rightarrow0$ and $J = 2\rightarrow1$ lines, respectively. 

These lines follow a similar 
spatial distribution and their respective peaks are located at about the offset position 
$\Delta\alpha=-120'', \Delta\delta=30''$, approximately 0.88 pc ($\sim80''$ at P.A. $90^{\circ}$) 
from the ridge. They both present extended emission, unlike a theoretically expected stratified PDR.
A spatial association between \thco\ and \ci\ was found by \citet{keene85}, on a scale size of 
$3'$ ($\sim2$~pc). We confirm these results with higher spatial resolution. We also add that, 
when looking at the overall distribution, the \co17\ and \ceio\ emissions are more similar to the \thco\ than to the \twco\ emission and, hence, they also show a spatial association with the \ci\ integrated emission. This spatial association is discussed further in the next section.

\subsection{The CO integrated intensity maps}

In order to compare them with the $J=1\to0$ transitions, and to increase the S/N of the $J=2\to1$ lines, all the maps 
(including the \ci\ lines from Fig.~\ref{fig:APEX-maps}) were convolved to the larger beam size (24$''$) of the 
\ceio~$J=1\to0$ line for the analysis presented in the next sections. Maps of the velocity integrated 
intensity of the \twco\ emission, and its isotopologues \thco, \co17, and \ceio\ lines, are shown in 
Figs.~\ref{fig:IRAM30m-maps1} and \ref{fig:IRAM30m-maps2}. 

\begin{figure}[!ht]

  \hfill\includegraphics[angle=0,width=0.24\textwidth]{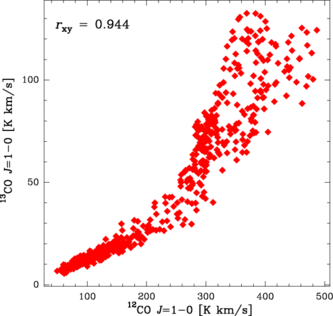}%
  \hfill\includegraphics[angle=0,width=0.24\textwidth]{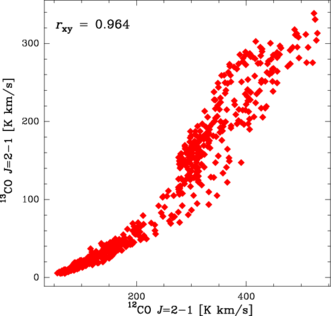}\hspace*{\fill}\\

  \hfill\includegraphics[angle=0,width=0.24\textwidth]{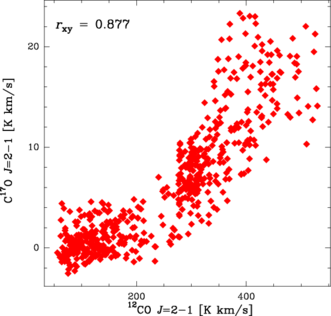}%
  \hfill\includegraphics[angle=0,width=0.24\textwidth]{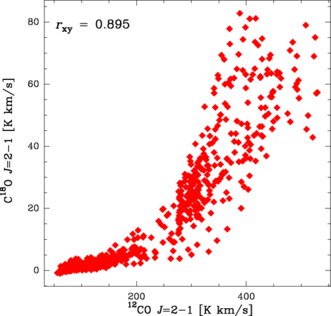}\hspace*{\fill}\\
  
  \hfill\includegraphics[angle=0,width=0.24\textwidth]{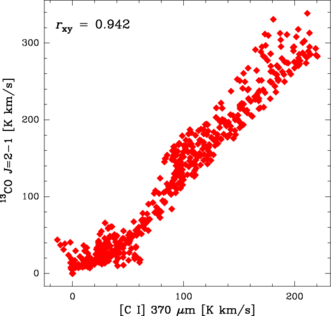}%
  \hfill\includegraphics[angle=0,width=0.24\textwidth]{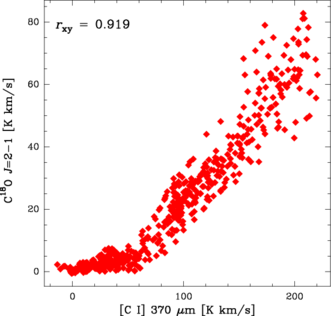}\hspace*{\fill}\\

  \caption{\footnotesize{Scatter plots and correlation coefficients $r_{xy}$ (from Eq.~\ref{eq:corr}) between the velocity-integrated intensity of \twco\ and \thco~$J=1\to0$ (\textit{top-left}), and the \twco~$J=2\to1$ compared to the $J=2\to1$ lines of \thco\ (\textit{top-right}), \co17\ (\textit{middle-left}), and \ceio\ (\textit{middle-right}). The \textit{bottom} panel shows the scatter plot and correlation coefficient between the \ci~370~\mum\ and the $J=2\to1$ transitions of \thco\ (\textit{bottom-left}) and \ceio\ (\textit{bottom-right}).
}}

  \label{fig:scatter-plots-check}
\end{figure}

The \twco\ integrated emission is  more extended, and its bulk emission does not resemble that of 
its isotope lines. In order to verify this, we use the scatter plots (Fig.~\ref{fig:scatter-plots-check}) of the 
velocity-integrated intensity of these tracers, and the corresponding correlation coefficient described in 
Appendix~\ref{sec:appendix-B}. The scatter plots deviate from the theoretical straight line expected for well-correlated maps. The \thco/\twco\ plots show a well-separated optically thin (low intensities) and an optically 
thick (high intensities) branch. It is interesting to note that the optically thick branch still has a 
relatively good correlation. This shows the integrated intensity growth via line broadening 
at lower densities, 
fully in line with Larson's law; i.e., the cloud size is inversely proportional to its density, and the velocity 
dispersion (or line broadening) is directly proportional to the cloud size, hence, the lower the density, the 
larger the cloud and the broader the lines.

We note, however, that even though the intensities of the \twco\ and \thco\ maps are very scattered 
(at the higher values) and show two branches with different slopes, the correlation coefficient is still 
relatively high, $r_{xy}=0.94-0.96$. These correlation coefficients are similar to those found 
between the $J=1\to0$ and $J=2\to1$ lines of \twco\ and \ci, which are much better correlated, as shown 
in Fig.~\ref{fig:scatter-area}. A similar case is found when comparing the \ci~370~\mum\ line with the 
$J=2\to1$ transitions of the \thco\ and \ceio\ lines (bottom panel in Fig.~\ref{fig:scatter-plots-check}). These 
scatter plots show values that are less scattered (supporting the similarity 
between the \ci\ and the CO isotopologue lines mentioned above), while the correlation coefficient is similar to  
(or lower than) those found for maps with \textit{\emph{less similar}} spatial distribution (e.g., \twco\ and 
\thco~$J=1\to0$). This means the correlation coefficient is not  robust enough to discriminate between ``well'' 
correlated and ``not so well'' correlated maps and, hence, it must be used with caution. Another way of using 
this statistical measure is to define a threshold to consider only regions of the map with high intensities, to 
compute the correlation coefficient as described in Sect.~\ref{sec:CII-correlation}.

Figure~\ref{fig:CO-chan-maps} shows the overlay between the 
two $J=1\to0$ and $2\to 1$ lowest energy 
transitions of \twco\ and the 
\cii~158~\mum\ line, both in the total velocity-integrated intensity and the 1~\kms channel maps. We note that the 
velocity-integrated intensity of \cii\ does not follow the spatial distribution of the \twco\ emission. However, 
some spatial association between these lines can be seen in the channel maps, but only at the central 
16--24~\kms velocity channels, where the bulk of the \twco\ emission is found.

New results in M17~SW have shown that 
a large fraction 
of the \cii\ emission is not associated with other species (e.g., \twco\ 
and \ci) tracing the star-forming material when analyzed at 
narrow 
velocity channels \citep{pb12, pb13}. 
Hence, line integrated intensity maps have to be interpreted with great care, as a smaller or larger part of the 
line integrated emission, as in the case of \cii\ here, may result in strong velocity components in one line, 
which are barely traceable in another line, like \ceio\ or \thco\ here, due to different physical origin and/or very 
different excitation conditions. This obviously  also strongly affects the interpretation of line-integrated 
intensity ratios between different tracers. 
In fact, we do not find strong spatial 
associations when comparing the velocity integrated \cii~158~\mum\ map from \pb\ \etal\ (2012, their Fig.~1) with 
the new CO and \ci\ maps from Figs.~\ref{fig:IRAM30m-maps1} and \ref{fig:IRAM30m-maps2}. Therefore, in the 
following sections we present our new analysis and discussions based on velocity channel maps, showing the 
intensities of the lines integrated over a narrow (1~\kms) channel width.

\section{\ Excitation and column density of \ci }

The critical density ($n_{cr}\sim10^3~\3cm$ for collisions with o-/p-H$_2$ at 100~K; from the 
LAMDA\footnote{http://www.strw.leidenuniv.nl/$\sim$moldata/} database, \citealt{schoier05}) and 
upper-level energy ($E_u\approx24$~K for $J=1\rightarrow0$, and $E_u\approx62$~K for 
$J=2\rightarrow1$) of \ci\  enable us to trace the diffuse ($n(\rm H_2)\leq10^3~\3cm$) ISM 
and estimate its temperature. We note, 
however, that this does not imply that \ci\ traces only diffuse gas, but also denser gas given the overall 
spatial association of the velocity-integrated intensities observed between \ci\ and the CO isotopologues, as 
described in section~\ref{sec:CI-results}. 

We first estimate the excitation temperature of \ci\ from the ratio between the two transitions, assuming 
optically thin emission. From this excitation temperature, the column density of \ci\ can be estimated as well, 
assuming optically thin emission and LTE conditions. Then we compare the results with a non-LTE estimate at 
representative offset positions. This is done in section~\ref{sec:CI-column}.

\subsection{ Excitation temperature of \ci }
\label{sec:CI-Tex}

We computed the ratio $R=I$(\ci~369\mum)/$I$(\ci~609\mum) between the intensities of 
the \ci\ lines integrated in narrow (1~\kms) velocity channels, for each channel map in the velocity range 
between 0~\kms\ and 40~\kms.
We 
find 
observed values of 1$\lesssim$\textit{R}$\lesssim $2, at the central 
velocity channels (16--24~\kms). 
Ratios lower than unity (which means subthermal excitation)
are also observed in our maps, but mostly at velocity channels $<$16~\kms\ and $>$24~\kms, where the intensity of 
both lines is below 30\% of their respective peak channel intensities, which is about the noise level of the 
fainter \ci~609\mum\ line.

The observed ratios match the values expected in a PDR environment with low 
density ($\la10^3~\3cm$) and relatively low radiation fields ($G_0\la10^2$, with $G_0$ in units of the ambient 
interstellar radiation field, $1.2\times10^{-4}~\rm ergs~s^{-1}~cm^{-1}~sr^{-1}$, \citealt{habing68}), as shown 
by \citet[][their Fig. 3]{meijerink07}. These densities are also in agreement with 
previous estimates by \citet[][]{meixner92}, who  concluded that the \cii, \ci,\ and low-$J$ CO lines emerge 
from an inter-clump medium, while the extended \cii\ and \ci\ emission emerges from a halo gas surrounding the 
clump and inter-clump material. However, the observed ratios are not exclusive of these ambient 
conditions, since they can also be found by extrapolating the values given by \citet[][their Fig. 3]{meijerink07}
at higher densities ($\sim10^4~\3cm$) and even lower radiation fields ($G_0\la10$), which are naturally 
attenuated by the larger column densities of denser clumps.

As mentioned in Sect.~\ref{sec:CI-results}, the total integrated intensity of the \ci\ emission resembles 
that of the optically thin isotope CO lines, rather than that of the \twco\ lines. The isotope lines have 
critical densities of $n_{crit}\sim10^4~\3cm$ (at $T_K~$100--200~K) and trace the compact emission from the 
denser molecular clouds. 
Because of the self-pumping due to large optical depths resulting from the high abundance, the \twco\ lines 
 trace not only the denser molecular clouds, but also the more diffuse and extended emission beyond (i.e., east 
from) the ionization front.
Hence, contrary to the general picture proposed by \citet{meixner92}, we favor a scenario where part of the \ci\ 
emission emerges from an inter-clump medium rather than from a more diffuse halo gas, and the other part is 
associated to the denser clumps traced by the isotope CO lines.

Following the formalism by \citet[][their Appendix A]{schneider03}, which is valid in the optically thin limit and  
  when both lines have similar optical depths, we can estimate the excitation temperature $T_{ex}$ of \ci\ 
from the ratio $R$ as $T_{ex}=38.8/ln(2.11/R)$~K. Figure~\ref{fig:CI-Tex} shows the estimated $T_{ex}$ of \ci\ 
at the same region mapped in \cii\, for the velocity channels from 9~\kms\ to 29~\kms where the emission of both \ci\ lines is larger than the $3\sigma$ level detection. The \ci\ maps were first 
convolved to the larger beam (24$''$) of the \ceio~$J=1\to0$ map in order to increase the S/N.

The excitation temperature ranges between $\sim$40~K and $\sim$100~K in the inner molecular region (i.e., southwest from the ionization 
front). This result is in agreement with a previous LTE estimate of the excitation 
temperature from mid-$J$ \twco\ \citep{pb10}, earlier estimates from \ci\ $J=2\rightarrow1$ observations 
\citep[e.g.,][]{genzel88} that indicated the \ci\ emission arises from gas with kinetic temperature of about 
50~K, and from a multi-line NH$_3$ study \citep{gusten88} which showed different coexisting gas phases with 
kinetic temperatures between 30~K and few 100~K, and up to about 275~K in the region traced by the VLA continuum 
arc (the northern ionization front). 
Even higher excitation temperatures are found at sparse locations northeast of the ionization front, 
and along the eastern edge of M17~SW at the channel bins (20--24~\kms) close to the 
ionization front. Pointing offsets between the CHAMP$^+$ and FLASH 
observations (done at different observing periods), as well as differential couplings of the two respective 
beams just at the edge of the ridge, can mimic such temperature gradients. These effects could account for up to 
30\% in the line ratios, and they cannot be discarded. 
Hence, we limit the color scale of Fig.~\ref{fig:CI-Tex} to 200~K, which is a more reliable upper limit of the 
excitation temperature in the region mapped. Ratios $>$2, found at few positions between the ionization 
front and the ionizing stars, lead to negative excitation temperatures in the optically thin and LTE 
approximation.

\begin{figure*}[!ht]

  \hfill\includegraphics[angle=0,width=0.7\textwidth]{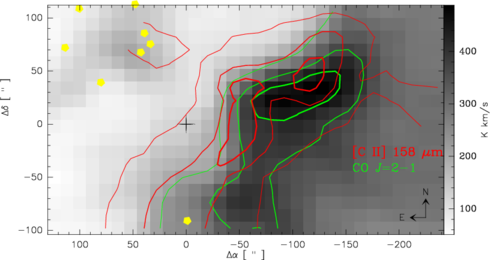}\hspace*{\fill}\\

  \hfill\includegraphics[angle=0,width=0.7\textwidth]{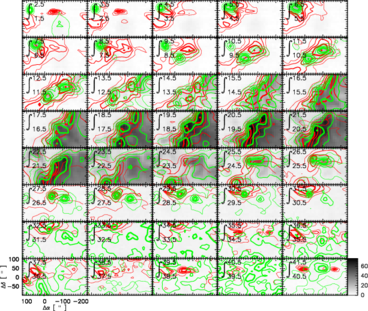}\hspace*{\fill}\\

  \caption{\footnotesize{\textit{Top} - Velocity-integrated intensity maps of \twco~$J=1\to0$ (gray), \cii~158~\mum\ (red contour), and \twco~$J=2\to1$ (green contour). The contour lines (from thin to thick) are  50\%, 75\%, and 90\% of the respective peak emissions. The \textit{stars} indicate the O and B ionizing stars \citep{beetz76, hanson97}. The reference position ($\Delta \alpha=0$, $\Delta \delta=0$), marked with a cross, is as in Fig.~\ref{fig:IRAM30m-maps1}. \textit{Bottom} - Velocity channel maps (at 1~\kms\ width) of the same lines as above. Contours are  20\%, 40\%, 60\%, 80\%, and 100\% of the respective peak emissions. All maps have been convolved with the largest beam of 24$''$ corresponding to the \ceio~$J=1\to0$ map. 
}}

  \label{fig:CO-chan-maps}
\end{figure*}

\begin{figure*}[!tp]

  \hfill\includegraphics[,angle=0,width=0.8\linewidth]{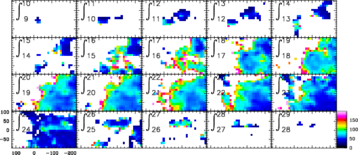}\hspace*{\fill}\\

  \caption{\footnotesize{Velocity channel maps (at 1~\kms\ width) of the \ci\ excitation temperature (color map in K) in M17~SW, estimated from the $R=I$(\ci~369\mum)/$I$(\ci~609\mum) line intensity ratio, and assuming LTE conditions and optically thin emission. Only velocity channels and pixels with emission larger than $3\sigma$ in both \ci\ lines are shown.
}}

  \label{fig:CI-Tex}
\end{figure*}

\begin{figure}[!htp]
 \begin{tabular}{c}
  \epsfig{file=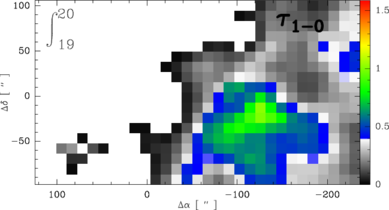,angle=0,width=0.85\linewidth}\\

  \epsfig{file=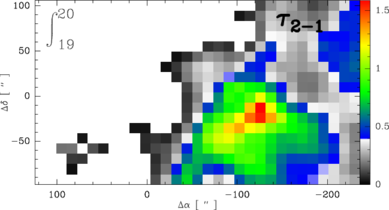,angle=0,width=0.85\linewidth}
  \end{tabular}

  \caption{\footnotesize{Optical depths of the \ci\ $J=1\to0$ (\textit{top}) and $J=2\to1$ (\textit{bottom}) 
  lines in the 19--20~\kms\ velocity channel map toward M17~SW (from Figs.~\ref{fig:CI-Tau10} and \ref{fig:CI-Tau21}), estimated from the excitation temperature of \ci, and assuming LTE conditions.
}}

  \label{fig:CI-Tau}
\end{figure}

\subsection{Optical depths and column density of \ci}
\label{sec:CI-column}

From the excitation temperature and the peak intensity of the \ci\ $J=1\rightarrow0$ and $J=2\rightarrow1$ 
lines, the optically thin approximation also allows us to estimate the optical depths of both lines. Knowing the 
excitation temperature and the optical depth of the $J=1\rightarrow0$ line, the column density $N($\ci$)$ can be 
computed as well. For detailed formulae see \citet{frerking89} and Schneider \etal\ (2003, their Appendix A). 
We also assume that the sources fully cover our beam in the emitting regions (i.e., we use a beam filling 
factor of unity). Therefore, the quantities reported here correspond to beam (24$''$) averaged values.

Figures~\ref{fig:CI-Tau10} and \ref{fig:CI-Tau21} show the channel maps of the optical depths estimated for the 
\ci\ $J=1\rightarrow0$ and $J=2\rightarrow1$ lines, respectively. Optical depths $\tau\leq1$ are 
observed in most of the regions mapped at all the velocity channels. Except in a small region around the 
offset position (-130$''$, -10$''$) of the $\tau_{2\to1}$ line, at the central velocity channels (19--20~\kms\ 
and 20--21~\kms), where the bulk of the \ci\ emission is found. 

A non-LTE excitation analysis using the Radex code \citep{vdtak07}, was used to test the optically thin 
assumption in two representative positions in the 19--20~\kms\ channel map. First at offset position 
(-130$''$, -10$''$) where $\tau_{2\to1}>1$ (cf. Fig.~\ref{fig:CI-xcmaps-pos1}), and then at the offset position 
(-130$''$, -70$''$) (cf. Fig.~\ref{fig:CI-xcmaps-pos2}) where $\tau_{2\to1}<1$, as seen in 
Fig.~\ref{fig:CI-Tau}.
The \ci\ line ratios and intensities observed at these positions can be reproduced with densities larger than 
$10^3~\3cm$ (the critical density of \ci~$J=1\to0$ at $T_K$ between 100~K and 200~K) for kinetic temperatures 
below 500~K, and column densities per line width $N/\Delta V=(4-7)\times10^{17}$ \ndv, similar to the 
column densities obtained with the LTE method. 
Considering a line width of $\sim$10~\kms, the column densities we obtained are consistent with the peak column 
density and a moderate optical depth of about 2.5$\pm$0.7 derived by \citet{genzel88} 
from the total velocity-integrated intensity of \ci~$J=2\to1$, and assuming a non-linear relation between the 
intensity of \ci\ and the intensity of \ceio. 

Our observed \ci\ intensities can also be reproduced with densities above $\sim$3$\times$10$^3~\3cm$ (the 
critical density of \ci~$J=2\to1$ at $T_k=$100--200~K) and temperatures below 300~K. The \ci~$J=1\to0$ line is 
close to thermalized ($T_{ex}\approx T_k$) at both positions, for $T_k<500$~K and densities 
$n(\rm H_2)=10^3-10^4~\3cm$, 
while $\tau$ is not much smaller than unity in both positions, just as in the LTE results. Although 
the optical depths of both \ci\ lines are just marginally thin, they are very similar. Hence, we can still apply 
the optically thin approximation in all the regions of interest, at least for $\tau_{1\to0}$, which is the 
opacity used to estimate the column density of \ci\ following \citet[][their Eq.~A.8]{schneider03}. Since the 
optical depth is not much smaller than unity, we used the correction factor $\tau(\ci)/(1-exp(-\tau(\ci)))$ to 
compute the column density of \ci.

The velocity channel maps of the column density $N($\ci$)$ ($\2cm$) are shown in Fig.~\ref{fig:CI-column}. The 
values of $N($\ci$)$ range between 10$^{15}~\2cm$ and $\sim$10$^{17}~\2cm$ throughout the whole 
region mapped and among all the velocity channels. However, the bulk of the \ci\ emission corresponds to column 
densities above $10^{16}\2cm$ in all the velocity channels. Column densities up to $10^{17}~\2cm$ are reached 
only in the central velocity channels, between 17~\kms\ and 22~\kms, which correspond to the regions where the 
integrated intensity maps (Fig.~\ref{fig:APEX-maps}) show a strong  \ci\ emission
 ($\ge50$\% of the peak).

\begin{figure*}[!hpt]

  \hfill\includegraphics[angle=0,width=0.8\textwidth]{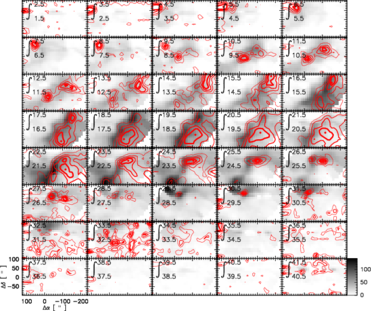}\hspace*{\fill}\\

  \caption{\footnotesize{Velocity channel map (integrated in 1~\kms) of the \cii~158\mum\ emission\ (gray background), with overlays of the $^3P_1\rightarrow~^3P_0$ fine-structure transition of \ci\ at 609\mum\ 
  (contours). The ionized and neutral carbon emissions are well correlated only at the intermediate velocities
  (16--24~\kms).
}}

  \label{fig:chan-map-CI1-0}
\end{figure*}

\begin{figure*}[!hpt]

  \hfill\includegraphics[,angle=0,width=0.8\textwidth]{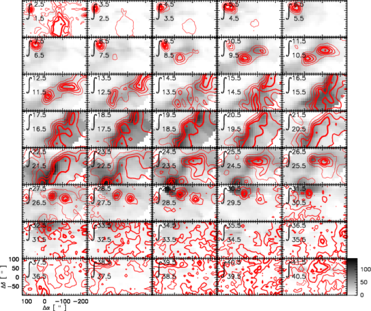}\hspace*{\fill}\\

  \caption{\footnotesize{Velocity channel map (integrated in 1~\kms) of the \cii~158\mum\ emission (gray background), with overlays of the \twco~$J=1\rightarrow0$ transition (contours). The molecular gas shows better 
  association with the ionized carbon at more extended intermediate velocities (14--27~\kms)  than the neutral carbon (Fig.~\ref{fig:chan-map-CI1-0}).
}}

  \label{fig:chan-map-CO1-0}
\end{figure*}

\begin{figure*}[!tp]

  \hfill\includegraphics[,angle=0,width=0.45\textwidth]{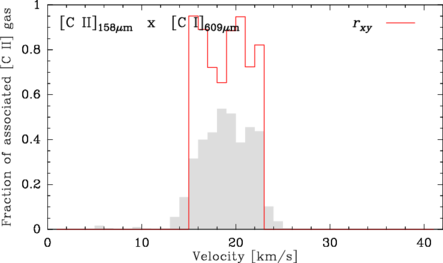}%
  \hfill\includegraphics[,angle=0,width=0.45\textwidth]{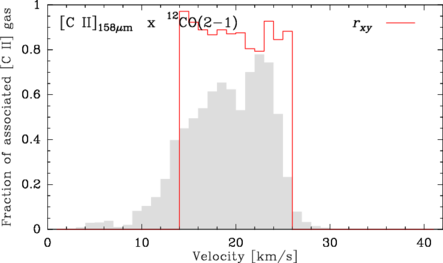}\hspace*{\fill}\\  

  \hfill\includegraphics[,angle=0,width=0.45\textwidth]{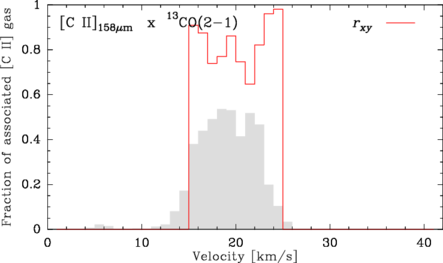}%
  \hfill\includegraphics[,angle=0,width=0.45\textwidth]{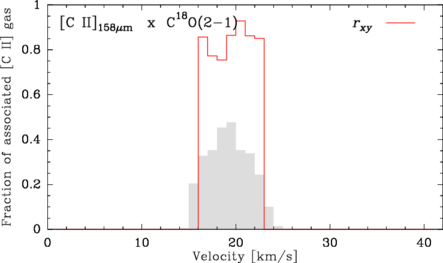}\hspace*{\fill}\\  

  \caption{\footnotesize{Histograms showing the fraction of the \cii~158~\mum\ emitting region correlated to each 1~\kms\ channel with other gas tracers. The corresponding correlation coefficient $r_{xy}$ is overlaid.
}}

  \label{fig:CII-correlations}
\end{figure*}

\section{\ Emission of \cii\  not associated with other gas tracers}
\label{sec:CII-emission}

In \pb\ \etal\ (2013) we showed overlays of our previous 
\twco\ $J=2\to1$ map and the optical depth of HI 
from \citet{brogan01}, over the \cii\ emission, also in channel maps of 1~\kms\ width. 
We showed that only at intermediate (10--24~\kms) velocities the \cii\ emission presented strong spatial 
association with other molecular gas tracers. 
A strong spatial association is identified when the spatial distribution of a particular \cii\ channel 
map is very similar, or adjacent, to that of another gas tracer (e.g., \twco~$J=2\to1$ or \ci).
On the other hand, at lower ($<$10~\kms) and higher ($>$24~\kms) velocity channels, the \cii\ emission is mostly 
not associated with the other tracers of diffuse and dense gas.
We note that ``not associated'' in this sense does not mean that we deal with physically completely independent 
material. The overlay between the \twco~$2\to1$ and \cii\ velocity channel maps in 
Fig.~\ref{fig:CO-chan-maps} shows 
that in the outer velocity range the \cii\ emission often shows halos and diffuse extensions around the denser 
clumps and filaments identifiable in the \twco\ channel maps, suggesting that the \cii\ emission traces gas that 
has been ablating off the clump or filament surfaces; this \cii\ emitting gas, however, is not visible in \twco, 
despite  this association. 

This can also be seen in the line shapes of the \cii, \ci~609~\mum, \ceio~$J=2\to1,$ and \twco~$J=1-0$ 
spectra shown in Fig.~\ref{fig:strip-line-spectra} for different 
offset positions along a strip line at position 
angle (P.A.)=63$^o$. The pointings of the spectra from the different tracer coincide within 2$''$, i.e., 
sufficiently close considering the smeared 24$''$ beam resolution.
We note that most of the lower and higher velocity channels of the \cii\ line are not 
associated with any of the other lines, while the \ceio\ and \ci\ lines are highly associated.
The velocity channel maps of \cii\ with overlays of the new maps of the \ci~609~\mum\ and \twco~$J=1\to0$ lines, 
are shown in Figs.~\ref{fig:chan-map-CI1-0} and \ref{fig:chan-map-CO1-0}, respectively. 

The spatial distribution of each velocity channel in the \ci~609~\mum\ emission follows a similar pattern to 
the maps presented in \citet{pb13}, although with a narrower (14--24~\kms) velocity range where strong 
association with the \cii\ emission is observed. 

If the \cii\ emission was associated in all velocity channels with the more diffuse molecular gas (between 
30 and $300~\3cm$, independent of temperature) traced by the \twco~$J=1\to0$ line, the spatial association  
between \cii\ and \twco~$J=1\to0$ would be expected to be stronger than that between \cii\ and \twco~$J=2\to1$.
However, like the   \twco~$J=2\to1$ and \ci\ lines, the \twco~$J=1\to0$ line shows a strong association with 
\cii\ \textit{\emph{only}} in the central 10-24~\kms\ components (cf. Fig.~\ref{fig:chan-map-CO1-0}).
This is another confirmation that the lower ($<$10~\kms) and higher ($>$24~\kms) velocity channels of 
the \cii\ emission are not strongly associated with the bulk of the molecular gas.

\subsection{Spatial correlation between the star-forming material and the \cii\ emission}
\label{sec:CII-correlation}

We quantify the spatial association observed at each velocity channel between the \cii\ emission and 
other gas tracers (e.g., \twco\ $J=1-0$, \ci~609~\mum) according to the procedure described in 
Appendix~\ref{sec:appendix-C}. We use 10\% of the global (i.e., among all the channels) peak emission of each tracer 
as a threshold to identify the pixels with significant emission to be used.
This method allows us to estimate \textit{\emph{where in the region}} and by \textit{\emph{what 
fraction of the region}} the \cii\ emission is associated, in the spatial distribution of each velocity channel map, with other gas tracers (cf. first paragraph in Sect.~\ref{sec:CII-emission}).

The correlation coefficient described in Appendix~\ref{sec:appendix-B}, and histograms showing the velocity 
distribution of the spatial association between the \cii\ emission and that of \ci~609~\mum, \twco, \thco, and 
\ceio~$J=2\to1$ are shown in Fig.~\ref{fig:CII-correlations}. In this case the correlation coefficient $r_{xy}$ 
was computed using only the pixels with emission larger than 1/3 of the global peak emission, in order to 
consider only the optically thick branch of the CO scatter plots shown in Fig.~\ref{fig:scatter-plots-check}, and 
to exclude the \ci\ scatter at low intensities.
 Figure~\ref{fig:CII-correlations} then shows that the \cii\ emission is associated with, and \textit{\emph{highly 
correlated with}} ($r_{xy}>0.6$)  other gas tracers, mostly at the central velocity channels between 15~\kms\ and 
23~\kms. The fraction of the mapped region where the \cii\ emission is associated with the \ci~609~\mum\ line is 
30\%--55\% in the velocity range mentioned above, while it reaches 40\%--80\% with \twco, 35\%--50\% with \thco, 
and only 20\%--45\% with \ceio~$J=2\to1$.

\begin{figure}[!pt]

 \begin{tabular}{c}

  \epsfig{file=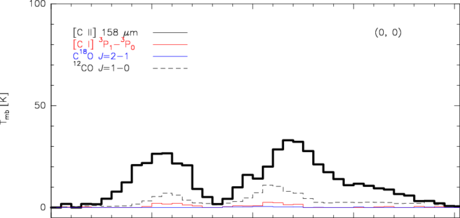,angle=0,width=0.98\linewidth}\\
  \vspace{-0.54cm}\\

  \epsfig{file=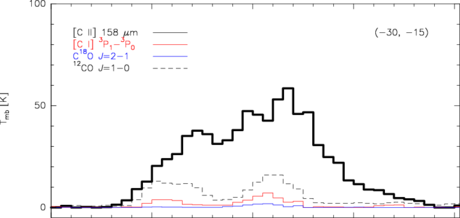,angle=0,width=0.98\linewidth}\\
  \vspace{-0.54cm}\\

  \epsfig{file=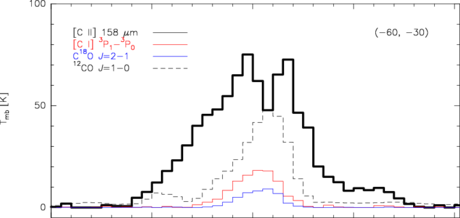,angle=0,width=0.98\linewidth}\\
  \vspace{-0.54cm}\\

  \epsfig{file=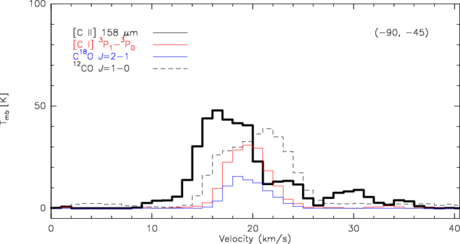,angle=0,width=0.98\linewidth}
 \end{tabular}

  \caption{\footnotesize{Spectra of several lines observed at approximated ($\pm$2$''$) offset positions along the strip line at P.A. $63^{\circ}$ ($\Delta\delta=\Delta\alpha/2$). All the spectra have been resampled to a 1~\kms\ resolution and convolved with the largest beam of 24$''$, corresponding to the \ceio~$J=1\to0$ map.}}

  \label{fig:strip-line-spectra}
\end{figure}

The large range of velocity channels of \cii\ emission not associated with other gas tracers is strong 
evidence of the inability of the total velocity-integrated \cii\ line intensity (i.e., its total flux)
to estimate its beam averaged 
abundance (i.e., column density ratio in comparison to other tracers),
and the cooling it provides to the molecular and star-forming gas traced by species like \ci\ and CO.
The assumption of the
\cii\ emission arising from the same spatial region as other tracers of diffuse and dense gas in  the whole 
velocity range covered by the \cii\ spectra, was the 
basis for many previous studies, 
according to the technology available at that time. In particular, the 
spectrometers 
on board of  NASA's Kuiper Airborne 
Observatory (KAO) had a spectral resolution of 80--175~\kms\ for the \cii~158~\mum\ line \citep[e.g.,][]
{meixner92}, while the \cii\ spans only $\sim$40~\kms\ in our velocity-resolved spectra. Now we know that only 
$\sim$20\% of that resolved velocity range is associated with the star-forming material traced by \ci\ and CO. 
Hence, using the total velocity-integrated \cii\ intensity can be misleading, since it may include (as in the 
case of M17~SW) emission from gas that is not really associated with the \ci\ or the CO emitting gas. Therefore, 
the actual abundance of C$^+$, and the cooling of the molecular gas due to \cii\ emission associated with 
molecular gas in several Galactic and extra-galactic environments, may be overestimated.

At the central velocity channels (e.g., 21--22~\kms) the \cii\ line shows dips at positions where optically thin 
lines (e.g., \ceio~$J=2\to1$) show their peaks, like the spectra in Fig.~\ref{fig:strip-line-spectra} at 
offset position (-60$''$, -30$''$). 
This spectral feature is present at several positions around the previous 
one, indicating the presence of a colder foreground layer, with 
significant optical depth to absorb the emission from a warmer background component present at an adjacent 
velocity. Taking the main beam temperature at the channel 22.5~\kms\ (where the \ceio~$J=2\to1$ lines drops 
sharply) as the continuum level ($T_{\rm C}\sim65.3$~K), and the temperature at the channel 21.5~\kms\ as 
the maximum absorption depth of the \cii\ line ($T_{\rm L}\sim47.4$~K), we can estimate the optical depth of the 
absorbing layer as $\tau=-ln(1-T_{\rm L}/T_{\rm C})$. Assuming that the absorbing layer completely covers the 
background component and that all the foreground \cii\ atoms are in the ground state,
we can estimate the absorbing column density following Eq.~(3) in \citet{neufeld10} as

\begin{equation}\label{eq:tau-CII}
\int \tau  {\rm d}v = \frac{A_{ul}g_{u}\lambda^3}{8 \pi g_{l}}N({\rm C^+}) = 7.2\times10^{-18}N({\rm C^+})~({\rm cm^{-2}})~{\rm km~s^{-1}}
,\end{equation}

\noindent
where $A_{ul}=2.3\times10^{-6}$~s$^{-1}$ is the spontaneous radiative decay rate, $g_u=4$ and $g_l=2$ 
are the degeneracies of the upper and lower states, and 
$\lambda=157.741$~\mum\ (used in cm in  Eq.~\ref{eq:tau-CII}) 
is the transition wavelength of the \cii\ ground state. From this we estimate a foreground absorbing column 
density of $N({\rm C^+})\approx2\times10^{17}~\2cm$. 
This is about a factor of two smaller than the 
column densities that we estimate  for the background emitting \cii\ gas (see Sect. 5.2). Since the 
absorbing layer affects only one or two of the central 1~\kms wide velocity channels, given the sharp profile of 
the \ceio\ line, it does not affect any of our conclusions.

In the following section we estimate in more detail  
the \cii\ column density and atomic hydrogen mass
not associated with the star-forming material traced by the \ci\ and \ceio\ lines by using an LTE approximation.

\subsection{Column density of \cii\ and mass of dissociated material}
\label{sec:CII-column-mass}

Because the spectra of \cii\ and the other gas tracers have slightly different velocity resolutions, we have 
re-sampled all the spectra to 1~\kms\ resolution.

We use the \ci~609~\mum\ line as a tracer of the diffuse ($n(\rm H_2)_{crit}\sim400~\3cm$ at 
$T_K\sim$100--200~K) molecular gas, and the optically thin \ceio~$J=2\to1$ line as a tracer 
of the denser ($n(\rm H_2)_{crit}\sim9\times10^3~\3cm$ at $T_K\sim$100--200~K) molecular gas in the star-forming 
material.

For each \ci\ and \ceio\ spectra of the map we find the channel with the maximum intensity, and divide 
the corresponding \cii\ channel by that maximum value to obtain the factors by which to multiply separately the 
\ci\ and \ceio\ spectra. This produced \textit{\emph{scaled up}} \ci\ and \ceio\ spectra that match the original \cii\ 
spectra at their respective maximum channel intensities. Then we subtract the scaled up \ci\ line from the 
original \cii\ spectra. The same is done independently with the \ceio\ line to produce two residual \cii\ 
spectra. Since we do not observe any absorption line profile in our spectra, we consider all the channels with 
negative values as noise, and we set them to zero in order to avoid unwanted boosting of the \cii\ emission.
If the maximum intensity of \ci\ or \ceio\ is higher than the intensity of the corresponding channel in 
the \cii\ spectrum, then no scaling up is done.
 
\begin{figure}[!tp]

 \begin{tabular}{c}

  \epsfig{file=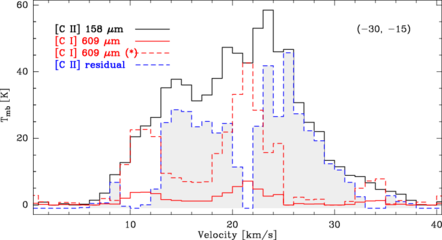,angle=0,width=0.9\linewidth}\\
  \vspace{-0.4cm}\\

  \epsfig{file=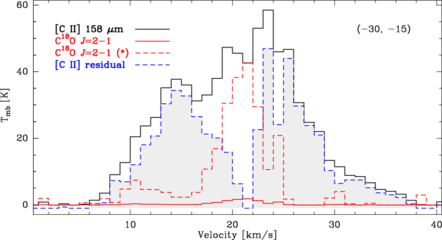,angle=0,width=0.9\linewidth}
 \end{tabular}
 \vspace{-0.25cm}\\

  \caption{\footnotesize{Residual \cii\ 158~\mum\ spectrum (dashed line and gray filled histogram) at offset 
  position (-30$''$,-15$''$) after subtracting the scaled up (marked with (*)) \ci~609~\mum\ (\textit{top}) 
  and \ceio~$J=2\to1$ (\textit{bottom}) lines from the original \cii\ spectrum. All negative (noise) channels in 
  the scaled up and residual spectra are set to zero. The residual \cii\ spectrum is shifted by -1 K for 
  clarity.}
}

  \label{fig:CII-residual-spec}
\end{figure}

An example of this procedure, and the results for the spectra at the approximated offset position 
(-30$''$,-15$''$), is shown in Fig.~\ref{fig:CII-residual-spec}. The shaded histogram corresponds to the 
residual spectra after the subtraction of the \ci\ and \ceio\ lines. For each channel we took the minimum 
intensity between the two residual \cii\ spectra to produce a synthetic spectrum that represents the \cii\ 
emission not associated with the star-forming material traced by \ci\ and \ceio. All these assumptions, in 
particular the scaling of the intensities, will lead to conservative numbers (lower limits).

We can estimate the column density $N(\rm C^+)$ of the non-associated \cii\ gas from the 
synthetic residual \cii\ spectra, following the high-temperature LTE limit, which is valid for temperatures well above 91~K and high densities,

\begin{equation}\label{eq:CII-column-LTE}
N({\rm C^+}) \approx \eta^{-1}I_{\rm [C II]} 6.3\times10^{20}~\2cm,
\end{equation}

\begin{figure}[!tp]

  \epsfig{file=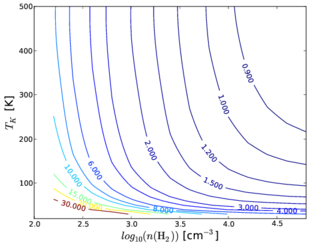,angle=0,width=0.9\linewidth}
  \epsfig{file=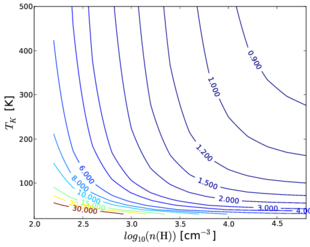,angle=0,width=0.9\linewidth}
  \epsfig{file=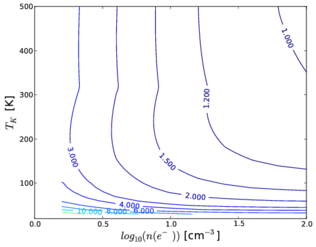,angle=0,width=0.9\linewidth}  
  \vspace{-0.25cm}
  
  \caption{\footnotesize{\cii\ column density enhancement for three collision partners, \hh\ (\textit{top}), \h1\ (\textit{middle}), and $e^-$ (\textit{bottom}), with respect to the column density obtained assuming a temperature of 250~K and a density of $10^4~\3cm$. This column density is depicted by the contour line equal unity.}
}

  \label{fig:CII-column-fraction}
\end{figure}

\noindent
obtained from a two-level system model as described in \citet[][their Eq.~(A.5),]{schneider03}
where $\eta_c$ is the beam filling factor assumed to be unity since the \cii\ emission in M17~SW is 
very extended, and $I_{\rm [C II]}$ is the \cii\ emission in units of erg cm$^{-2}$ s$^{-1}$ sr$^{-1}$.
We note that with the two-level system expression, the estimated column density of \cii\ increases if lower 
densities and/or lower temperatures are used. 
If we assume a gas temperature of 250~K and a density of $10^4~\3cm$, as 
estimated for the region east of the ionization front in M17~SW (i.e., the \h1\ and \hii\ regions) from 
mid-$J$ \twco\ line observations (\pb\ \etal\ 2010), the exact expression for the two-level system 
presented by \citet[][their Eq.~(A.4),]{schneider03} gives a 25\% larger column density than the LTE 
approximation of Eq.~(\ref{eq:CII-column-LTE}). 
Since the procedure described above subtracts the \cii\ emission associated with most of the dense and 
diffuse molecular gas, the residual emission should be dominated by collisional excitation from atomic 
hydrogen and free electrons.

In order to analyze how sensitive the \cii\ column density is to an assumed temperature 
and density of the gas, we estimate $N(\rm C^+)$ for a range of temperatures and densities using Eq.~(A.4) from \citet{schneider03}, considering a filling factor $\eta_c$ of unity. We consider the critical densities $n_{cr}$ for free electrons ($e^-$), atomic hydrogen (\h1), and molecular hydrogen (\hh), which were computed for each temperature, according to the corresponding collisional 
deexcitation rate coefficients reported by \citet{barinovs05}.
Since the actual column density of \cii\ also depends on the \cii\ intensity (which is arbitrary for this 
analysis), in Fig.~\ref{fig:CII-column-fraction} we show only the ratio with respect to the $N(\rm C^+)$ obtained using the temperature (250~K) and density ($n(\rm H)=10^4~\3cm$) assumed above, to demonstrate the relative effect of using different ambient conditions. Although the critical density depends on the temperature, 
$n_{cr}$ changes by a small percent between 200~K and 500~K. Therefore, we adopt 250~K as a high-temperature limit, since the temperature dependence of the exponential term in the two-level system approximation of $N(\rm C^+)$ is stronger.

 When using a higher temperature and density, the estimated column density decreases by less than 20\% when 
considering \h1\ or \hh\ as the collision partner. 
The column density of \cii\ would increase by larger factors if lower densities and temperatures 
were used. When using electrons as the collision partners, $N(\rm C^+)$ saturates at densities above 
$100~\3cm$, and temperatures above 400~K. These results emphasize our point that the values of $N(\rm C^+)$ 
obtained with the LTE approximation (or with the temperature and density assumed above) should be regarded as 
lower limits, whether the LTE conditions are met or not in all the regions mapped.

\begin{figure*}[!tp]

  \hfill\includegraphics[angle=0,width=0.7\textwidth]{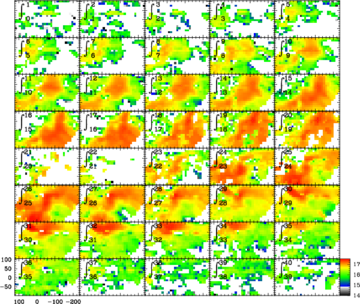}\hspace*{\fill}\\

  \caption{\footnotesize{Velocity channel maps at 1~\kms\ width of the residual \cii\ column density ($\2cm$ in $log_{10}$ scale), estimated assuming LTE conditions, that is not associated with the dense and halo molecular gas traced by \ceio\ and \ci, respectively.}
}

  \label{fig:CII-column}
\end{figure*}

The channel maps of the residual \cii\ column density (not associated with the dense and halo molecular 
gas traced by \ceio\ and \ci), estimated with Eq.~(\ref{eq:CII-column-LTE}), is shown in 
Fig.~\ref{fig:CII-column}.
The channel maps between 19~\kms\ and 22~\kms\ are the most affected by the subtraction of the \ci\ and \ceio\ 
emission, thus confirming the strong spatial association found in the central velocity range, as shown in 
Sect.~\ref{sec:CII-correlation}. The small self-absorption that we see will affect the neighboring 
channels, but not the general picture. This is in line with the channel maps from 18--20~\kms\ that still follow 
the structure of M17SW.
The column density of the \cii\ gas not associated with the star-forming 
material (traced by \ci\ and \ceio) ranges between $\sim$10$^{14}~\2cm$ and 
$\sim$4$\times$10$^{17}~\2cm$ in the whole region mapped, and among all the channels in the {0--40}~\kms\ 
velocity range.

From the residual (and total) \cii\ column density channel maps we can also compute the 
corresponding mass of the gas contained in the 24$''$ beam area of the channel maps by assuming a 
gas phase carbon abundance of $X(\rm{C^+/H})=1.2\times10^{-4}$ \citep[][their Table~1]{wakelam08} and complete ionization of the carbon, according to

\begin{equation}\label{eq:HI-mass}
M_{\rm gas} = 1.4 m_{\rm H} \frac{ N({\cii}) }{X_{\rm C^+/H}} A_{beam}
,\end{equation}

\noindent
where $A_{beam}$ is the area (in cm$^2$) covered by the 24$''$ beam, $m_{\rm H}$ is the atomic hydrogen 
mass (in g), and the factor 1.4 accounts for helium and a minor fraction of other heavier elements.

\begin{figure}[!tp]

 \begin{tabular}{c}
  \vspace{-0.65cm}\\
  \epsfig{file=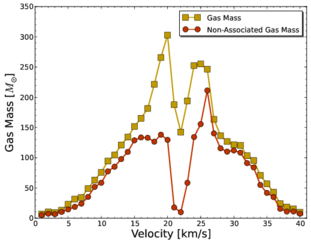,angle=0,width=0.85\linewidth}\\
  \vspace{-0.4cm}\\

 \end{tabular}
 \vspace{-0.25cm}\\

  \caption{\footnotesize{Gas mass (squares) estimated from the \cii~158~\mum\ emission with 
  Eq.~(\ref{eq:HI-mass}) at each velocity channel in the range 0~\kms\ to 40~\kms. The gas mass not associated 
  in the spatial distribution of each velocity channel map with star-forming material traced by the 
  \ceio~$J=2\to1$ and \ci~609~\mum\ lines is shown by filled circles.}
}

  \label{fig:Mass}
\end{figure}

The velocity distribution of the mass (obtained by adding up all the pixels from each velocity channel map 
created from Eq.~\ref{eq:HI-mass}) is shown in Fig.~\ref{fig:Mass}. The gas mass per velocity channel 
(squares) was estimated from the original \cii\ spectra. 
The largest fraction of non-associated gas mass (circles) is found at the higher (25--33~\kms) velocity 
channels.

When integrating the mass in the 0--40~\kms\ velocity range, we find a gas mass of 
$\sim$4.4$\times$10$^3$~\Msun\ in the entire region mapped. This mass is a factor $\sim$3 lower than the 
1.45$\times$10$^4$~\Msun\ found by \citep{stutzki90} from \ceio\ observations, which trace the non-dissociated 
cloud core mass. 
We note that the mass obtained from the LTE estimated column density of \ci\ 
(cf. Fig.~\ref{fig:CI-column}) using the same Eq.~\ref{eq:HI-mass} is about three orders of magnitude lower 
than the mass traced by \cii. This is due to the lower intensity and less spatial extension of the \ci\ emission 
throughout the mapped region and among all the velocity bins, compared to that of \cii. This is in line with the 
fact that only a small fraction of carbon is expected to be in atomic gas traced by the \ci\ line.

Considering the mass estimated from the residual \cii\ spectra, the gas mass from the non-associated \cii\ emission is $\sim$2.8$\times$10$^3$~\Msun. Thus, the estimated gas mass 
not associated with the star-forming material traced by \ci\ and \ceio\  corresponds to $\sim$64\% of the total gas mass traced by the original \cii\ emission. This still 
amounts to at least 19\% of the \ceio\ mass reported by \citet{stutzki90}.

A source of uncertainty to consider in our analysis is that we are assuming that the volumes of gas
corresponding to each velocity channel are associated. In other words, a spatial association of the \cii\ 
emission with \ci\ and \ceio\ emitting regions in the plane of the sky, does not ensure that they are really 
associated along the line of sight. However, the probability that the spatial association that we see in the central 
components of the 1~\kms\ channel maps is just a projection effect is minimal since we are using a relatively 
narrow velocity width. Hence, the fraction of non-associated mass quoted above should be considered a 
lower limit.

Since there is no evidence of fast shocks in M17~SW, other mechanisms that can produce \cii\ emission with 
larger velocity dispersion than the denser molecular gas must be considered. For instance, the interaction with 
winds and outflows from the ionizing stars can lead to substantial excitation of the \cii\ emitting gas. Hence, 
ablation \cite[e.g.,][and references therein]{castor75, weaver77, tenorio79, henley12} and, probably slow shock-
interaction due to radiative pressure \citep[e.g.,][and references therein]{goodwin97, krumholz10, dale11}, have 
to be considered to model and interpret the \cii\ emission not associated with the star-forming material in 
M17~SW. However, \cii\ may also be present in extended low density gas around \hii\ regions produced by far 
ultraviolet (FUV) photons (\textit{E}$>$11.26~eV for the first ionizing potential of atomic 
carbon) from the ionizing stars. 
With such high energy photons, low \hh\ densities also mean higher abundance (i.e., densities) of \h1\ and free 
electrons $e^-$, which would then become equally important collision partners of \cii, with critical densities of 
about 3$\times$10$^3~\3cm$ and 10~$\3cm$, respectively, at 250~K (the higher the temperature, the higher the 
critical densities). Thus, compensating the lower densities of \hh\ and, most likely, keeping the LTE 
assumption for \cii\ valid. Therefore, warm C$^+$  could be present at lower densities (and higher gas 
temperatures) than 
assumed. If so, then extreme ultraviolet (EUV) photons should also produce \nii\ emission if their energy is 
larger than 14.5~eV (the first ionizing potential of nitrogen) and lower than 24.38~eV (the second 
ionizing potential of carbon). In such a zone of EUV photons, the \cii\ and \nii\ emission should 
co-exist, and show a high degree of spatial association. This can be checked observationally, using the GREAT 
instrument on board  SOFIA. Zones with photon energies larger than 24.59~eV (the first ionizing potential of 
helium) accounts for at least  10\% of the gas in 
M17~SW (depending on the spatial resolution of the
observations) as estimated from observations of the He$^+$/H ratio \citep[e.g.,][]{peimbert88, tsivilev99}.
Between the two extremes of molecular and ionized hydrogen, the \cii\ emission can also co-exist with 
neutral atomic hydrogen \h1, as shown by \citep{brogan01} in M17~SW. We discuss   these gas phases in more detail in 
the following section.

\subsection{\cii\ in the three gas phases }
\label{sec:CII-gas-phases}

\begin{figure}[!tp]

 \begin{tabular}{c}

  \epsfig{file=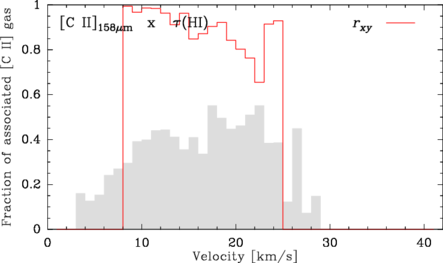,angle=0,width=0.9\linewidth}\\
 \end{tabular}
 \vspace{-0.25cm}\\

  \caption{\footnotesize{Fraction of the \cii~158~\mum\ emitting region correlated (at each 1~\kms\ channel) with the optical depth of \h1\ \citep[from][]{brogan01}. The corresponding correlation coefficient $r_{xy}$ is overlaid.}
}

  \label{fig:CII-tauHI-correlation}
\end{figure}

\begin{figure}[!tp]

 \begin{tabular}{c}

  \epsfig{file=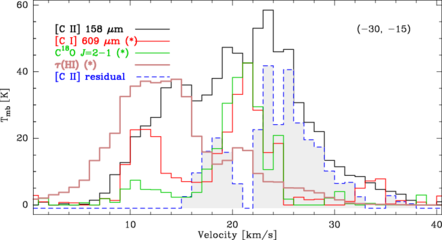,angle=0,width=0.9\linewidth}\\
  \epsfig{file=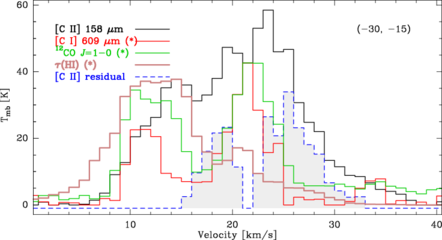,angle=0,width=0.9\linewidth}\\
  \epsfig{file=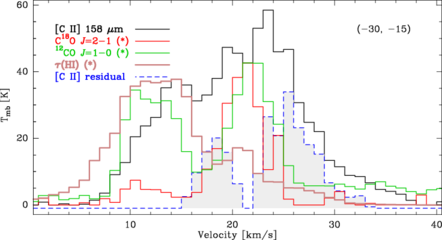,angle=0,width=0.9\linewidth}\\  
 \end{tabular}
 \vspace{-0.25cm}\\

  \caption{\footnotesize{Residual \cii~158~\mum\ spectrum (dashed line and gray filled histogram) at offset 
  position (-30$''$,-15$''$), after subtracting, from the original \cii\ spectrum, the scaled up (*) spectra of 
  model 1-(top), 2-(middle) and 3-(bottom), as well as the optical depth of \h1, $\tau$(\h1) (see text). 
  All negative (noise) channels in the scaled up and residual spectra are set to zero. The residual \cii\ 
  spectrum is shifted in -1 K for clarity.}
}

  \label{fig:CII-tauHI-residual-spec}
\end{figure}

As discussed above, there are basically three different regimes that contribute to the \cii\ emission; 
the highly ionized gas where electrons dominate (\hii), the atomic hydrogen layer (\h1), and the molecular 
hydrogen gas (\hh) suffused with sufficient UV to keep CO dissociated and to ionize neutral carbon 
efficiently\footnote{There is also the CO-dark molecular gas \citep{wolfire10} that we  trace in part 
through our \ci\ line. However our \ci\ sensitivity is not good enough to also measure the more diffuse 
molecular, CO-dark gas.}.
Following the method described above, we include the high resolution VLA map of the velocity-resolved 
optical depth, $\tau$(\h1), from \citet{brogan01}, convolved with a 24$''$ beam and re-sampled to 1~\kms\ 
channel width. 
The channel-by-channel spatial correlation between \cii\ emission and $\tau$(\h1) shown in 
Fig.~\ref{fig:CII-tauHI-residual-spec} indicates that most of the \cii\ emission associated 
with the \h1\ gas is found at the lower ($<20$~\kms) velocity channels.

In the previous section we estimated the \cii\ column density and hydrogen mass of the gas not 
associated with the relatively compact and dense star-forming material traced by \ceio~$J=2\to1$, and \ci~609~\mum. The \cii\ emission associated with the entire molecular gas phase, 
however, also comprises  the diffuse and more extended \hh\ gas. Therefore, we now use the \twco~$J=1\to0$ line, 
as the canonical tracer of \hh. As \twco~$J=1\to0$ is optically thick throughout large parts of the map, this will provide a lower limit for the \cii\ emission from the diffuse molecular material.

\begin{figure*}[!t]

 \begin{tabular}{ccc}
  \vspace{-0.65cm}\\

  \epsfig{file=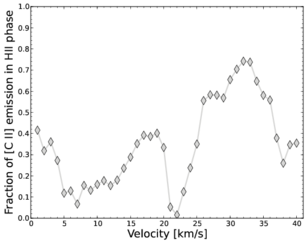,angle=0,width=0.32\linewidth} &
  \hspace{-0.4cm}\epsfig{file=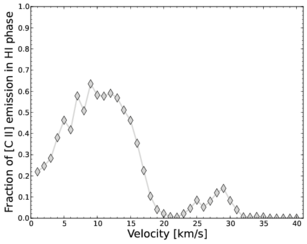,angle=0,width=0.32\linewidth} &
  \hspace{-0.4cm}\epsfig{file=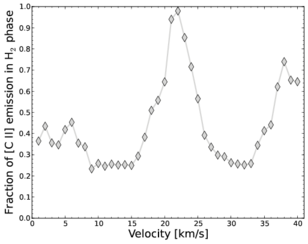,angle=0,width=0.32\linewidth}    
 \end{tabular}
 \vspace{-0.25cm}\\

  \caption{\footnotesize{Fraction of the average over the region mapped residual \cii\ emission corresponding to the three gas phases: \hii\ (\textit{left}), \h1\ (\textit{middle}), and \hh\ (\textit{right}) as obtained with model (2) in Table~\ref{tab:CII-phases}.}
}

  \label{fig:CII-phases}
\end{figure*}

In order to analyze the impact of these line tracers in the residual \cii\ emission (and hence, the \cii\ column and 
associated gas mass), we tested three different combinations of gas tracers: (1) $\tau$(\h1) + \ci~609~\mum\ + 
\ceio(2--1); (2) $\tau$(\h1) + \ci~609~\mum\ + \twco(1--0); and (3) $\tau$(\h1) + \twco(1--0) + \ceio(2--1).
Model (3) is included for comparison, since \ceio\ can complement \twco\ in regions where \twco\ is optically 
thick.
The residual \cii\ spectrum at offset position (-30$''$,-15$''$), obtained after subtracting the scaled-up 
spectra of the three combinations mentioned above, are shown in Fig.~\ref{fig:CII-tauHI-residual-spec} 
(\textit{\emph{from top to bottom}}). When subtracting (channel by channel) the maximum of the three synthetic lines 
from the original \cii\ spectra, we obtain the residual \cii\ emission that is mostly associated with the \hii\ 
regime. It can be seen  in the three cases that most of the residual \cii\ emission is contained in the higher 
velocity channels.
Using only the molecular gas tracers (i.e., excluding $\tau$(\h1)), we can obtain a second residual 
\cii\ spectra that would contain the \cii\ emission associated mostly with the \hii\ and \h1\ gas. From 
these two residual spectra we can then estimate the residual \cii\ emissions, and their respective column 
densities using Eq.~(\ref{eq:CII-column-LTE}), associated with the three gas phases, according to the 
procedure described in Appendix~\ref{sec:appendix-D}.

In Table~\ref{tab:CII-phases} we summarize the fraction of the average over the region mapped \cii\ 
emission associated with the three gas phases as obtained from the three combinations of line tracers. 
Using \ci~609~\mum\ and \twco~$J=1\to0$, combined with $\tau$(\h1), yields 
practically the same result given the uncertainties as when using \ceio~$J=2\to1$ instead of \ci~609~\mum. 
This results from the high correlation observed between \ci\ and \ceio, although it is not a 1:1 match (probably 
sensitivity limit driven) as shown in Fig.~\ref{fig:scatter-plots-check}. Thus, the slightly higher ($\sim$1\%) 
fraction of \cii\ emission associated with \hh\ gas found with model (2), compared with that of model 
(3), may indicate that \ci\ traces at least part of the CO-dark molecular gas \citep{wolfire10}. Therefore, we 
chose model (2) as the most complete one, tracing all the gas regimes where \cii\ emission can be found.

\begin{table}[!pt]
 \caption[]{Fraction of average \cii\ emission in the three gas phases.}
 \label{tab:CII-phases}
 \centering
 \begin{tabular}{lccc}
   \hline\hline
   \noalign{\smallskip}
   Model No. & \hii\ fraction & \h1\ fraction    & \hh\ fraction \\
   \noalign{\smallskip}
   \hline
   \noalign{\smallskip}
     
      Model 1\tablefootmark{a}   &   42.7\%   &   20.9\%   &   36.4\%  \\
      Model 2\tablefootmark{b}   &   36.2\%   &   16.8\%   &   47.0\%  \\
      Model 3\tablefootmark{c}   &   36.8\%   &   17.4\%   &   45.8\%  \\

   \noalign{\smallskip}
   \hline                           
 \end{tabular}
\tablefoot{
\tablefoottext{a}{Combination: $\cii-\{\tau(\h1)+\ci_{609~\mum}+\ceio~(2-1)\}$}
\tablefoottext{b}{Combination: $\cii-\{\tau(\h1)+\ci_{609~\mum}+\twco~(1-0)\}$}
\tablefoottext{c}{Combination: $\cii-\{\tau(\h1)+\twco~(1-0)+\ceio~(2-1)\}$}
}
\end{table}

The velocity distribution of the fraction of averaged \cii\ 
emission associated with the three gas phases obtained with model (2) is shown in 
Fig.~\ref{fig:CII-phases}. The \cii\ emission associated with \h1\ gas is mostly contained in the lower velocity ($<$20~\kms) channels, while the \cii\ emission associated with the ionized \hii\ gas is contained mainly at the higher 
velocity bins ($>$25~\kms),  although part of it is also found in the $<$20~\kms\ velocity range. The central velocity channels (15--30~\kms) contain most of the \cii\ emission associated with 
the molecular \hh\ gas.
The corresponding velocity-channel maps of the \cii\ emission associated with \hii, \h1, and \hh, are 
shown in Fig.~\ref{fig:CII-H2mass-phases}. These channel maps show that the \cii\ emission, and therefore column 
density, associated with the ionized gas, peaks at the northeast corner of the mapped region, which 
coincides with the position of the ionizing sources (cf. Fig~\ref{fig:CO-chan-maps}).
The fraction of \cii\ column density (or \cii\ emission) associated with the molecular gas regime is 
about 11\% larger than the fraction (36\%) found in the dense star-forming material. 
This is expected since the 
\twco\ emission has a broader line profile and is also spatially more extended than \ceio\ and \ci\ (cf. 
Fig.~\ref{fig:IRAM30m-maps1} and Fig.~\ref{fig:strip-line-spectra}).

We note that this method has uncertainties, and it gives only a first order approximation of 
the \cii\ emission associated with the three different regimes. The 
results presented in Table~\ref{tab:CII-phases} should not be taken as a sharp distinction between the three gas 
regimes, since in reality the three gas phases can be mixed throughout the region mapped (we 
elaborate  on this in the next sections).
In particular, the \cii\ emission associated with the atomic \h1\ gas has a large uncertainty 
as the optical depth $\tau$(\h1) is saturated over a significant part of the region along the molecular ridge. 
The saturated values were replaced by a lower limit, according to the continuum and rms level of the VLA spectra, 
as described in detail by \citet[][their Sect.~3.3.2]{brogan01}. Furthermore, $\tau$(\h1) obtained from \h1\ 
in absorption traces atomic hydrogen in the foreground relative to the free-free emission from the \hii\ 
region. This might introduce a bias since $\tau$(\h1) traces only the cold \h1\ gas, while part of the warmer 
atomic hydrogen can be mixed with the \hh\ and \hii\ gas phases. However, we consider that the warm mixed
\h1\ gas can be at least partially accounted for by the \ci\ and \twco\ lines. Nevertheless, this is another 
uncertainty in our method.

The gas masses associated with the three gas phases could be estimated with Eq.~(\ref{eq:HI-mass}), by using 
abundances (and the corresponding mass of atomic and molecular hydrogen) of ionized carbon relative to the 
dominant hydrogen phase, i.e., $X(\rm C^+/H^0)$, $X(\rm C^+/H_2)$. 
However, these abundances are not really known, and although they could be estimated from a clumpy PDR 
model, which would be the best model currently available  for M17~SW because of its highly clumpy structure, the 
uncertainties in the values obtained for each gas phase would be very high  because the relative 
abundances strongly depend on the number of clumps, clump sizes, and ambient conditions of each clump, which we 
have not yet been able to constrain for M17~SW. In addition, knowing the actual density and temperature of the dominant 
collision partners in each gas regime would allow us to estimate (with non-LTE radiative transfer models) the 
actual \cii\ emission associated with the three gas phases. We expect to estimate all these parameters in a 
follow up work.

\begin{figure*}[!ht]

 \begin{tabular}{cc}
  \hfill\includegraphics[angle=0,width=0.45\textwidth]{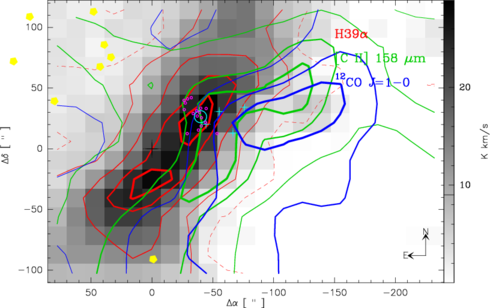}\hspace*{\fill} &
  \hfill\includegraphics[angle=0,width=0.45\textwidth]{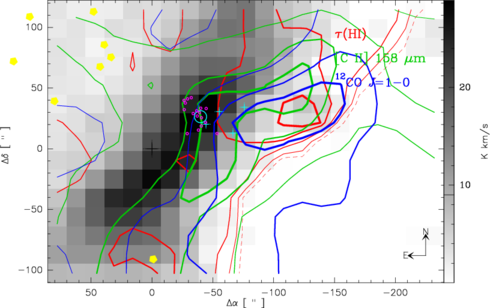}\hspace*{\fill}
 \end{tabular}
 
  \caption{\footnotesize{\textit{Left} - Velocity-integrated intensity maps of H41$\alpha$ (gray), H39$\alpha$ (red contour), \cii~158~\mum\ (green contour), and \twco~$J=1\to0$ (blue contour). The contour lines (from thin to thick) are  10\% (dashed line), 25\%, 50\%, 75\%, and 90\% of the respective peak emissions. The \textit{stars} indicate the O and B ionizing stars \citep{beetz76, hanson97}. The reference position ($\Delta \alpha=0$, $\Delta \delta=0$), marked with a cross, is as in Fig.~\ref{fig:IRAM30m-maps1}. The ultracompact \hii\ region M17-UC1 and four H$_2$O masers \citep{johnson98} are marked by the circle and ``+'' symbols, respectively. The \textit{\emph{small purple circles}} correspond to the heavily obscured ($E_{median}>2.5$ keV, $A_V\ge10$ mag) population of X-ray sources around the M17-UC1 region (Fig.10 in Broos~\etal\ 2007; coordinates from the VizieR catalog).
  \textit{Right} - Same as on the left, but with $\tau$(\h1) instead of H39$\alpha$. All maps have been convolved with a 30$''$ beam, to increase the S/N of the H41$\alpha$ map. 
}}

  \label{fig:RRL-overlay-maps}
\end{figure*}

\subsection{Comparison with radio recombination lines}
\label{sec:RRL}

Traditionally radio recombination lines (RRL) are considered probes of gas conditions within ionized gas. 
However, the principal quantum levels are generally not in thermal equilibrium, and there are different competing 
mechanisms governing the line intensities (e.g., maser amplification of background radiation and weakening due to 
underpopulation of the upper quantum levels relative to LTE conditions) as well as the broadening of the line 
widths (e.g., thermal, turbulence and Stark effects - \citealt{griem67}). Furthermore, the line shapes are the result 
of emission from regions with different ambient conditions along the line of sight, since the emitting gas is 
usually optically thin in the centimeter and millimeter regimes. All these effects make of RRLs, in the 
centimeter wavelength range, more ambiguous probes of the ambient conditions in ionized gas than originally 
thought. At millimeter wavelengths, however, the Stark broadening is 
negligible ($\Delta V\sim c(\frac{\lambda}{100\ {\rm m}})^{5/3}\leq$0.1~\kms\ for $\nu\geq90$~GHz, cf. 
\citealt{gordon02}), leaving only thermal and turbulent broadening at play, and the RRLs tend to be optically 
thin and do not suffer significant departures from LTE conditions (for a review see, e.g., \citealt{gordon88, gordon02}).

In our broadband (32 GHz IF bandwidth) OTF maps obtained with the IRAM 30m telescope (cf. 
Sect.~\ref{sec:observations}), we detect a number of hydrogen RRLs in the 3mm band: H39$\alpha$, H40$\alpha,$ and 
H41$\alpha$ lines at 106.74 GHz, 99.02~GHz, and 92.03~GHz, respectively (the beam size of the H41$\alpha$ line is 
28$\farcs$3). All of them show very similar spatial distribution and line shapes. We improved the S/N by 
convolving the maps with a 30$''$ beam. 
Because of our short integration times no other recombination lines were detected, preventing further analysis of the 
excitation conditions of the ionized gas at our high spatial resolution. 
Our goal now is simply to verify if the spatial distribution of the \cii\ emission can be associated with the 
ionized gas traced by the hydrogen recombination lines.

Figure~\ref{fig:RRL-overlay-maps} shows the H39$\alpha$, \cii,\ and \twco~$J=1\to0$ integrated intensities, 
as well as the $\tau$(\h1) optical depth overlay (right panel)  on the H41$\alpha$ map. The hydrogen 
recombination lines follow a relatively homogeneous distribution about 45$^{\circ}$ along the ionization front, 
and they peak at the M17-UC1 ultracompact region, which is also surrounded by a number of embedded X-ray sources 
\citep{broos07}. We note that there are many more X-ray sources reported by Broos~\etal,  several of them 
with a stellar counterpart, but for clarity of the figures we show here only those associated with the M17-UC1 
region.
The \h1\ optical depth follows the distribution of the RRLs, but their peak emissions are not correlated. In 
fact, the velocity integrated $\tau$(\h1) peaks around the same position as the \twco~$J=1\to0$ line intensity. 
The \cii\ emission, instead, covers the entire region mapped, and its intensity peaks between the ionized gas 
traced by H41$\alpha$ and the molecular gas traced by the \twco~$J=1\to0$ line.


\begin{figure}[!pt]

 \begin{tabular}{l}
  \epsfig{file=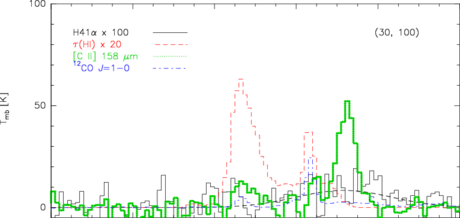,angle=0,width=0.98\linewidth}\\
  \vspace{-0.54cm}\\

  \epsfig{file=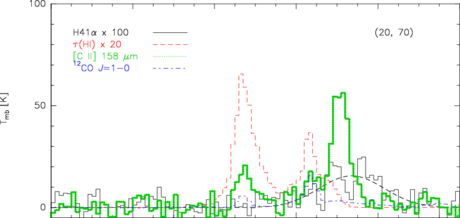,angle=0,width=0.98\linewidth}\\
  \vspace{-0.54cm}\\

  \epsfig{file=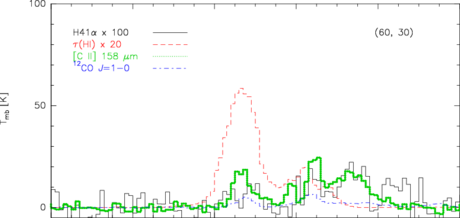,angle=0,width=0.98\linewidth}\\
  \vspace{-0.54cm}\\

  \hspace{0.0cm}\epsfig{file=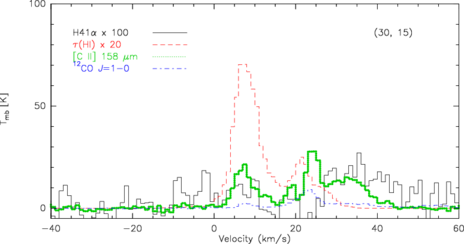,angle=0,width=0.99\linewidth}

 \end{tabular}

  \caption{\footnotesize{Spectra of several lines observed at approximated ($\pm$3$''$) offset positions along the strip line at P.A. $63^{\circ}$ ($\Delta\delta=\Delta\alpha/2$) and at positions close to the ionizing stars. All the spectra have been resampled to a 1~\kms\ resolution and convolved with a 30$''$ beam, to increase the S/N of the H41$\alpha$ map. A Gaussian was fit to the H41$\alpha$ line at the top two positions.}}

  \label{fig:RRL-spectra1}
\end{figure}

\begin{figure}[!pt]

 \begin{tabular}{l}

  \epsfig{file=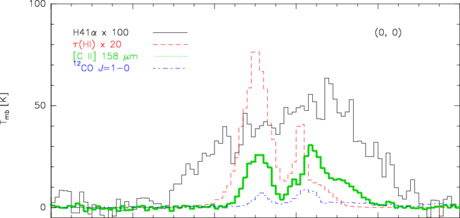,angle=0,width=0.98\linewidth}\\ 
  \vspace{-0.54cm}\\
  
  \epsfig{file=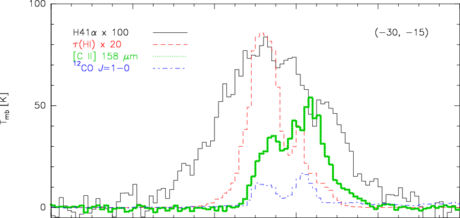,angle=0,width=0.98\linewidth}\\ 
  \vspace{-0.54cm}\\  
  
  \epsfig{file=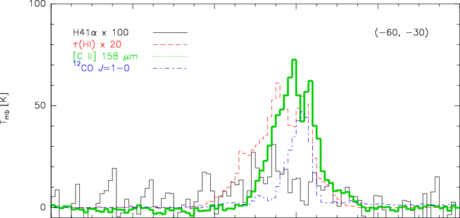,angle=0,width=0.98\linewidth}\\ 
  \vspace{-0.54cm}\\  
  
 \hspace{0.0cm}\epsfig{file=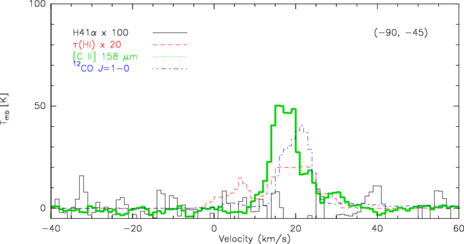,angle=0,width=0.99\linewidth}

 \end{tabular}

  \caption{\footnotesize{Spectra of several lines observed at approximated ($\pm$3$''$) offset positions along the strip line at P.A. $63^{\circ}$ ($\Delta\delta=\Delta\alpha/2$). All the spectra have been resampled to a 1~\kms\ resolution and convolved with a 30$''$ beam, to increase the S/N of the H41$\alpha$ map.}}

  \label{fig:RRL-spectra2}
\end{figure}

Figures~\ref{fig:RRL-spectra1} and \ref{fig:RRL-spectra2} show the spectra of the H41$\alpha$, \cii,\ and \twco~$J=1\to0$ lines at two offset positions close to the ionizing sources, (30$''$,100$''$) and (20$''$,70$''$), and along a strip line at P.A.=63$^{\circ}$. Although fainter at the northeast region (Fig.~\ref{fig:RRL-spectra1}), the H41$\alpha$ line shows emission only at the higher velocity range ($>$20~\kms) coinciding with the velocity channels of the \cii\ lines found to be associated mostly with the \hii\ gas in Sect.~\ref{sec:CII-gas-phases}. At the southwest region (Fig.~\ref{fig:RRL-spectra2}) the H41$\alpha$ line is stronger, but its shape is asymmetric and much broader than any of the other lines we observed in M17~SW. It is hard to associate its line shape to any of the line structures observed in the \cii\ line nor in $\tau$(\h1). Since the H41$\alpha$ line is expected to be optically thin (i.e., no optical depth nor self-absorption effects), its line shape is most likely formed by several layers of ionized gas (with different $V_{lsr}$) along the line of sight, with each velocity component being affected by a combination of thermal and turbulence broadening.

\begin{figure}[!pt]

 \begin{tabular}{cc}

  \epsfig{file=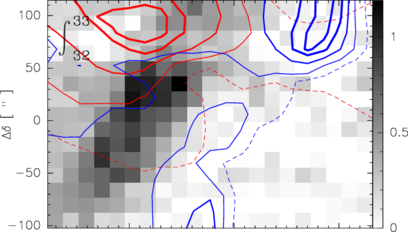,angle=0,width=0.45\linewidth} & 
  \epsfig{file=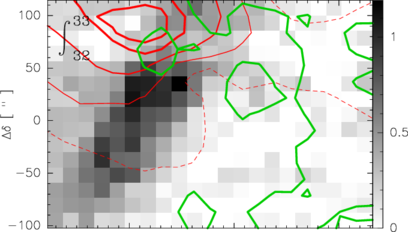,angle=0,width=0.45\linewidth}\\
  \vspace{-0.54cm}\\
  
  \epsfig{file=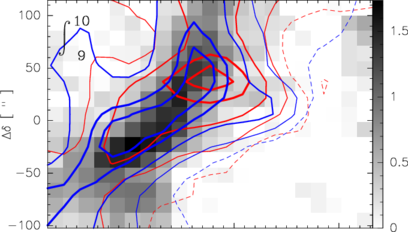,angle=0,width=0.45\linewidth} &
  \epsfig{file=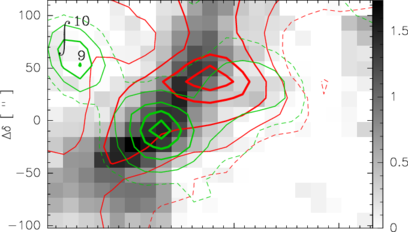,angle=0,width=0.45\linewidth}\\   
  \vspace{-0.54cm}\\  
  
  \epsfig{file=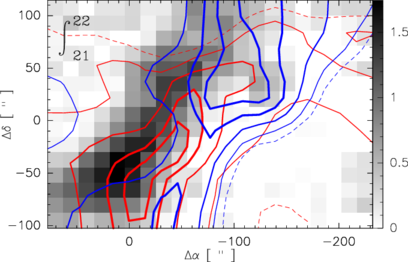,angle=0,width=0.45\linewidth} &
  \epsfig{file=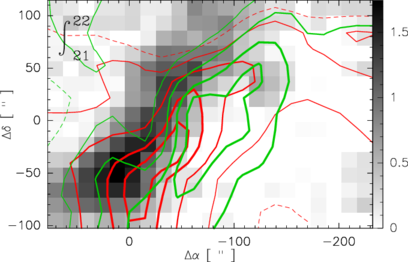,angle=0,width=0.45\linewidth}\\   
    
 \end{tabular}

  \caption{\footnotesize{Channel maps of the H41$\alpha$ line (gray background, in \Kkms) with overlays of the \cii\ (red contours), the $\tau$(\h1) optical depth (blue contours) (\textit{left panels}), and the \twco~$J=1\to0$ (green contours) (\textit{right panels}), at the velocity ranges of the peak \cii\ emission associated with (from top to bottom) the \hii, \h1, and \hh\ gas regimes. The corresponding velocity ranges are shown in the top-left of each channel map. The contour lines (from thin to thick) are  10\% (dashed line), 25\%, 50\%, 75\%, and 90\% of the respective peak emissions. All the maps were convolved with a 30$''$ beam to increase the S/N of the H41$\alpha$ map.}}

  \label{fig:RRL-chan-maps}
\end{figure}

From Fig.~\ref{fig:CII-phases}, we know that the velocity ranges of the peak \cii\ (residual) emission found to 
be associated with the \hii, \h1, and \hh\ gas, are 32--33~\kms, 9--10~\kms, and 21-22~\kms, respectively.
Figure~\ref{fig:RRL-chan-maps} shows the corresponding channel maps of H41$\alpha$, the original \cii\ emission, 
$\tau$(\h1), and the \twco~$J=1\to0$ line. We note that the spatial distribution of the H41$\alpha$ line does not 
change significantly over these three different velocity ranges. At 32--33~\kms, the \cii\ emission peaks at the 
northeast region, around the ionizing sources, while the H41$\alpha$ emission still peaks at the M17-UC1 region 
(cf. Fig.~\ref{fig:RRL-overlay-maps}). At 21-22~\kms, the H41$\alpha$ peak is at the southeast region of the 
ionization front, while the distribution of the \cii\ emission follows closely that of the \twco~$J=1\to0$. In 
the velocity range 9--10~\kms, identified mostly with the \h1\ gas regime, the emission from all the lines seems 
to emerge mostly from the region of the ionization front traced by the H41$\alpha$ line, indicating that the 
three gas regimes must be mixed at these velocities. The fact that the ionization and molecular dissociation 
front are mixed, can be explained if advection is considered \citep[e.g.,][]{bertoldi96}. This dynamical
process would make the ionized region larger compared to the atomic region, and extended towards the molecular 
region. This, in turn, will lead to a larger contribution of the ionized gas to PDR diagnostics like \cii, as 
shown in models by \citet{abel05}.

As previously noted by \citet{brogan99}, and from Figs.~\ref{fig:RRL-spectra1} and \ref{fig:RRL-spectra2}, 
the velocity structures of \h1\ and \cii\ are different. A narrow component of shocked \h1\ gas (streaming 
towards us) in M17 was suggested by \citet{brogan99}, in the range 11--17~\kms, where we also find \cii\ and 
\twco\ components, but only at the edge of the ionization front (cf. Fig.~\ref{fig:RRL-spectra2}, offset 
position ($-30''$,$-15''$)).
In the 9--10~\kms\ velocity range, which is associated mostly with atomic gas, the \h1\ optical depth 
peaks along the H41$\alpha$ line since $\tau$(\h1) is estimated from \H1\ in absorption. Actually, this channel 
map is part of the 
prominent $\tau$(\h1) velocity component in the 0--11~\kms\ range (cf. Fig.~\ref{fig:RRL-spectra1}), that has 
corresponding associated \cii\ and \twco~$J=1\to0$ emission. This component in $\tau$(\h1) was not mentioned in 
the original papers by \citet{brogan99,brogan01}. This can also be shocked ionized (\cii), atomic (\h1), 
and molecular (CO) gas streaming towards us.

The strong \cii\ velocity component found around $V_{lsr}=32$~\kms\ (Fig.~\ref{fig:RRL-spectra1}, top two panels) 
has no evident counterpart in the other atomic or molecular tracers. These differences, in particular with \h1, 
may arise because the \cii~158~\mum\ emission tends to trace warmer gas than the \h1\ absorption,  which traces 
cold gas. The faint H41$\alpha$ emission around $V_{lsr}=32$~\kms, barely detected with S/N$\sim$3 in our OTF maps at offset position (20$''$,70$''$), shows a line width (FWHM) of 
$\Delta V=20\pm3$~\kms\ from a Gaussian fit. This line width is consistent with the thermal broadening 
($\Delta v_{G-thermal}\sim c \times 7.16233\times 10^{-7} ( \frac{T}{M} )^{1/2}$, with $T$ in K and $M$ in  
 amu, cf. \citealt[][their Eq.~2.22]{gordon02}) expected for the LTE electron temperature 
$T_e^*=10700\pm700$~K estimated towards M17 by \citet{gordon89}. 
At offset (30$''$,100$''$) we find a S/N$<$3 for the H41$\alpha$ line.
At these positions, (20$''$,70$''$) and (30$''$,100$''$), the \cii\ component around 
$V_{lsr}=32$~\kms\ have a Gaussian fit FWHM of $\sim5.5\pm0.5$~\kms, also consistent with the thermal 
broadening expected for the $T_e^*$ quoted above. This indicates that the \cii\ emission around 
$V_{lsr}=32$~\kms\ is associated with 
warm gas, and probably shocked by the proximity of the ionizing sources, streaming away from us and from the 
bulk of the molecular unshocked gas found around the 20~\kms\ component. Higher S/N maps of the H41$\alpha$ and 
other RRLs are needed  to confirm the detection of ionized gas at these positions.  
Maps of \nii\ and the hydrogen recombination lines (in the $\geq$3~mm wavelength range), tracing the transitions 
between medium and large principal quantum numbers, and even the fainter carbon recombination lines, may help
explain the nature and ambient conditions of the 32~\kms\ component found in the \cii\ emission.

\subsection{Comparison with other sources and implications for extragalactic studies}

A study of the \hii\ region S125 showed that up to 40\% of the \cii~158~\mum\ line intensity, observed with 
ISO-SWL, arises from the \hii\ region \citep{aannestad03}.
Comparing and modeling the \cii\ and \nii\ emission obtained from Herschel/PACS, \citet{bernards12} found that 
at most 18\% of the \cii\ emission in the Orion Bar  originated in the \hii\ region. This is a factor of $\sim$2 
lower than found for S125, and a factor of $\sim$ 3.6 times lower than what we estimate for M17~SW. 
However, this is not a one-to-one comparison, since their study is 
focused in a relatively small scale, covering mostly the dense PDR in the Orion Bar. Hence, most of the actual 
\hii\ region (where more \cii\ emission than quoted by them can be present) is not seen in their data set. 
Therefore, large scale maps of \cii\ and other ionized, neutral, and molecular tracers like the ones we present are relevant not only in order to understand the full picture of the \hii-PDR boundary, which  is more complex than 
predicted by the classical 1-D models, but also to understand and properly interpret the observations of 
these diagnostic lines in extragalactic sources.

As mentioned at the end of Sect.~\ref{sec:CII-column-mass}, and as shown in 
Sect.~\ref{sec:RRL}, the \cii\ line can originate both in the 
PDR (molecular and neutral atomic gas) and in the \hii\ region.
When observed toward other galaxies, the \cii\ emission collected in one beam (typically covering large scales), 
will naturally come mainly from several unresolved giant molecular clouds and \hii\ regions, with different 
orientations with respect to the impinging radiation field along the line of sight. Thus, it is
important to characterize in detail the \cii\ contribution from these different environments.
The \cii\ emission is important in extragalactic studies for redshift determinations, and also 
to estimate the star formation rate (SFR) from the \cii\ luminosity \citep[e.g.,][]{stacey91, meijerink07, 
luhman03}.

Recent results from 130 galaxies observed with Herschel/PACS by \citet{sargsyan14} confirm that \cii\ 
traces the same starburst component of sources as measured with mid-infrared PAH, neon emission line diagnostics, 
and bolometric luminosities of reradiating dust from starbursts.
By using a modified Cloudy code \citet{abel05} estimated that about 30\% of the \cii\ emission in the 
starburst galaxy NGC~253, observed with the KAO \citep{carral94}, arise from the ionized medium. 
Using PACS data to map the FIR 
emission of the \cii\ line in the spiral galaxy M33, and models of photo-ionization and photon-dominated regions, 
\citet{mookerjea11} found that between 20\%\ and 30\% of this emission comes from the \hii\ region.

Compared  with our lower limit estimates, these results for extragalactic sources 
are at least a factor of two lower than the values we find for M17~SW when considering only the \cii\ emission not 
associated with the dense star-forming material (cf. Sect.~\ref{sec:CII-column-mass}). They are more than 6\% 
lower than what we estimate for the \cii\ emission associated mostly with the \hii\ region (cf.
Sect.~\ref{sec:CII-gas-phases}). We believe that there are three main reasons for these differences:
\begin{itemize}
\item[(1)] Their models of static geometries do not consider that their beams collect emission from many sources with diverse geometries (from edge-on to face-on, and many other orientations in between), systemic velocities, highly structured medium (that would allow FUV radiation to permeate the region ionizing and heating the gas on larger spatial scales), and dynamical processes like advection (as mentioned in Sect.~\ref{sec:RRL}) that would increase the contribution from the ionized medium.
\item[(2)] The models assume that the gas is in pressure equilibrium, so 
the equation of state is dominated by gas pressure since turbulent and magnetic pressures 
are not included. If turbulent or magnetic pressure dominate instead, as is the case of M17~SW 
\citep{pellegrini07}, then the density law and hence the collisional excitation of the PDR 
diagnostic lines throughout the \hii\ and photon-dominated regions will be different.
\item[(3)] Their studies are based on the total velocity-integrated intensities of the \cii\ and other line 
emissions, provided by the spectral resolution of the instruments used. Whereas the velocity-resolved \cii\ 
spectra obtained with SOFIA/GREAT towards M17~SW, shows that $\sim$80\% of the velocity range covered by 
the \cii\ line is associated with a mixture of ionized medium, neutral gas, and diffuse molecular gas, and only 
$\sim$20\% of the velocity range is associated mostly with the dense (star-forming) molecular material (cf. 
Sects.~\ref{sec:CII-correlation} to \ref{sec:RRL}). While this result is particular for M17~SW, we have 
found similar results in other regions (e.g., NGC~3603, \citealt{pb15}), showing that this characteristic should 
be taken into account in PDR modeling of Galactic and extragalactic sources.
\end{itemize}

All the reasons mentioned above lead us to think that the fraction of \cii\ emission coming from the 
\hii\ regions have been underestimated in previous studies. Their conclusions regarding the fraction of \cii\ 
emission emerging from ongoing star formation, and their estimates of SFRs in comparison with other SFR 
diagnostics, may change. In order to judge how much more valid our estimate is, models of photo-ionization 
and photon-dominated regions that include dynamical processes and magnetic pressure are needed. In addition, a 
significant number of large scale maps of \cii and other diagnostics of ionized gas toward Galactic sources are 
required to increase the statistical significance of our results. 
Large scale maps of many sources can be achieved with the upGREAT receiver arrays, the second 
generation receivers for the GREAT project, that will make mapping the large scale regions more efficient. The 
upGREAT low frequency array (1.9--2.5 THz with 14 pixels) is currently scheduled for commissioning onboard 
SOFIA in May 2015. The upGREAT high frequency array (4.7 THz with 7 pixels) will follow about one year later. 
We expect to show the results of this long term study in follow up work.

\section{Conclusions}

We used the dual channel DSB receiver FLASH on the APEX telescope
to map (with $12\farcs 7$ resolution) a region of about 4.1 pc $\times$ 4.7 pc in the $^3P_1\rightarrow{^3P_0}$ 609 
$\mu$m ($J=1\rightarrow0$) fine-structure transition of \ci, towards the star-forming region M17~SW. We also 
used the broadband EMIR receivers on the IRAM 30m telescope to map a similar area of 360$''\times$300$''$ in 
the 3mm, 2mm, and 1mm bands.

We combine these data with the previously observed and published \ci~$^3P_2\rightarrow{^3P_1}$~370~\mum\ and 
\cii~158~\mum\ data to discuss the physical properties of the dense interstellar medium in the M17~SW ridge.

Because of the complex structure of the M17~SW and the availability of velocity resolved spectra, we have 
performed all our analysis on 1~\kms-wide velocity channel maps.

\subsection{\ Excitation and column densities of \ci}

Combining our earlier observation of the \ci~$^3P_2\rightarrow{^3P_1}$~370~\mum\ ($J=2\rightarrow1$) fine-structure line together with the new \ci~$^3P_1\rightarrow{^3P_0}$~609~\mum\ data, 
we found that the $R=I$(\ci~369\mum)/$I$(\ci~609\mum) ratio is larger than unity in most of the 
regions mapped, at the central (10--24~\kms) velocity channels where the bulk ($>$20\%) of the \ci\ emission is 
found. We estimate the excitation temperature $T_{ex}$ and column density of \ci\ using an optically thin 
approximation and a non-LTE method. We found that $T_{ex}$ ranges between $\sim$40 K and $\sim$100 K in the 
inner region (i.e., southwest from the ionization front).
While the \ci\ column density ranges between $\sim$4$\times$10$^{13}~\2cm$ and $\sim$10$^{17}~\2cm$ throughout 
the whole region mapped and among all the velocity channels. In the region where the \ci\ emission is $\ge50$\% 
of its peak integrated intensity, column densities up to $10^{17}~\2cm$ are only reached  in the central 
(17-22~\kms) velocity channels.

\subsection{Comparison of \cii, \ci,\ and CO}

We used our recent SOFIA/GREAT velocity-resolved \cii\ map to analyze the spatial and velocity association of 
the \cii\ emission with the diffuse and molecular gas in M17~SW. For that we used the \ci\ lines from APEX/FLASH 
and the \twco\ and isotope lines from IRAM30m/EMIR. 

The \cii\ emission was found to be associated with the other gas tracers in 20\%--80\% of the mapped 
region, but only at the central (15~\kms\ and 23~\kms) velocity channels, which means that only $\sim$20\% of the velocity range ($\sim$40~\kms) that the \cii\ line spans in our velocity-resolved spectra is associated with the star-forming material in M17~SW.

For the non-associated \cii\ gas,  in the 1~\kms\ wide channel maps we estimated column densities ranging from 
$\sim$10$^{14}~\ndv$ to $\sim$5$\times$10$^{17}~\ndv$ across the region mapped. 
The largest fraction of non-associated atomic mass contained in the mapped 
region, is found at the higher (25--33~\kms) velocity channels.
The total non-associated gas mass (integrated over the 0--40~\kms\ range) is 
$\sim$2.8$\times$10$^3$~\Msun. This corresponds to a very large fraction ($\sim$64\%)  of the 
total mass ($\sim$4.4$\times$10$^3$~\Msun) traced by the \cii\ emission. In other words, most of the gas 
traced by the \cii\ emission in M17~SW is not associated with the star-forming material. The good match 
of the \ci\ emission with the \ceio~$J=2\to1$ map indicates that both tracers are seen from the same, star-forming material.

\subsection{Comparison of \cii, \h1,\ and hydrogen recombination lines}

When using the optical depth of \h1 in combination with the \ci~609~\mum\ and 
\twco~$J=1\to0$ maps, we found that the \cii\ emission associated with \h1\ is contained in the lower 
($<$20~\kms) velocity channels, while the \cii\ emission associated with the \hii\ gas phase is found 
mostly at the higher velocity ($>$25~\kms) channels. Most of the \cii\ emission 
associated with the diffuse and dense molecular \hh\ gas, is found in the central velocity channels 
(15--30~\kms). From our preferred model (2), we found that 36.2\%, 
16.8\%, and 47.0\% of the \cii\ emission is contained in the \hii, \h1, and \hh\ regimes, 
respectively.

Overlays between \cii, $\tau$(\h1), \twco~$J=1\to0$, and the H41$\alpha$ line at velocity ranges associated with 
the three gas regimes indicate that the \hii\ region is mixed with the atomic and part of the molecular 
dissociation regions, in agreement with the highly clumped structure and dynamical processes at play in M17~SW.

Considering M17~SW as a proxy for active galaxies, these results are also relevant to extra-galactic 
studies in which \cii\ is often used as a tracer of star formation rates.
We have estimated a fraction of the \cii\ emission not associated with the dense star-forming material, based on 
large maps of velocity-resolved spectra. Our estimates are up to two times larger than estimates done in 
extragalactic sources, based on velocity-integrated \cii\ intensities.
Other Galactic molecular clouds, for which velocity-resolved \cii\ maps are also available, will be analyzed in a 
follow-up work. With the enhanced observing capabilities of SOFIA's second generation heterodyne arrays, more 
sources can be observed in the future, increasing the statistics.

\begin{acknowledgements}
We are grateful to the MPIfR team, as well as the
APEX and IRAM 30m staff for their help and support during and after the observations.
We are grateful to C. Brogan for providing the 21~cm map and the data cube of the velocity-resolved optical depth 
of HI estimated for M17~SW.
We thank the referee for the careful reading of the manuscript and constructive comments that helped to improve 
our work. We are also grateful with the editor, Dr. Malcolm Walmsley, for his timely comments and suggestions that helped to improve even more our work.
Molecular Databases that have been helpful include the NASA/JPL spectroscopy line catalog and the University of 
Leiden's LAMDA databases.
\end{acknowledgements}


\bibliographystyle{aa}
\setlength{\bibsep}{-2.1pt}
\bibliography{m17sw}

\begin{thebibliography}{91}
\expandafter\ifx\csname natexlab\endcsname\relax\def\natexlab#1{#1}\fi

\bibitem[{{Aannestad} \& {Emery}(2003)}]{aannestad03}
{Aannestad}, P.~A. \& {Emery}, R.~J. 2003, \aap, 406, 155

\bibitem[{{Abel} {et~al.}(2005){Abel}, {Ferland}, {Shaw}, \& {van
  Hoof}}]{abel05}
{Abel}, N.~P., {Ferland}, G.~J., {Shaw}, G., \& {van Hoof}, P.~A.~M. 2005,
  \apjs, 161, 65

\bibitem[{{Banerjee} {et~al.}(2004){Banerjee}, {Pudritz}, \&
  {Holmes}}]{banerjee04}
{Banerjee}, R., {Pudritz}, R.~E., \& {Holmes}, L. 2004, \mnras, 355, 248

\bibitem[{{Barinovs} {et~al.}(2005){Barinovs}, {van Hemert}, {Krems}, \&
  {Dalgarno}}]{barinovs05}
{Barinovs}, {\u G}., {van Hemert}, M.~C., {Krems}, R., \& {Dalgarno}, A. 2005,
  \apj, 620, 537

\bibitem[{{Beetz} {et~al.}(1976){Beetz}, {Elsaesser}, {Weinberger}, \&
  {Poulakos}}]{beetz76}
{Beetz}, M., {Elsaesser}, H., {Weinberger}, R., \& {Poulakos}, C. 1976, \aap,
  50, 41

\bibitem[{{Bernard-Salas} {et~al.}(2012){Bernard-Salas}, {Habart}, {Arab},
  {Abergel}, {Dartois}, {Martin}, {Bontemps}, {Joblin}, {White}, {Bernard}, \&
  {Naylor}}]{bernards12}
{Bernard-Salas}, J., {Habart}, E., {Arab}, H., {et~al.} 2012, \aap, 538, A37

\bibitem[{{Bertoldi} \& {Draine}(1996)}]{bertoldi96}
{Bertoldi}, F. \& {Draine}, B.~T. 1996, \apj, 458, 222

\bibitem[{{Brogan} {et~al.}(1999){Brogan}, {Troland}, {Roberts}, \&
  {Crutcher}}]{brogan99}
{Brogan}, C., {Troland}, T., {Roberts}, D., \& {Crutcher}, R. 1999, \apj, 515,
  304

\bibitem[{{Brogan} \& {Troland}(2001)}]{brogan01}
{Brogan}, C.~L. \& {Troland}, T.~H. 2001, \apj, 560, 821

\bibitem[{{Broos} {et~al.}(2007){Broos}, {Feigelson}, {Townsley}, {Getman},
  {Wang}, {Garmire}, {Jiang}, \& {Tsuboi}}]{broos07}
{Broos}, P.~S., {Feigelson}, E.~D., {Townsley}, L.~K., {et~al.} 2007, \apjs,
  169, 353

\bibitem[{{Carr}(1987)}]{carr87}
{Carr}, J.~S. 1987, \apj, 323, 170

\bibitem[{{Carral} {et~al.}(1994){Carral}, {Hollenbach}, {Lord}, {Colgan},
  {Haas}, {Rubin}, \& {Erickson}}]{carral94}
{Carral}, P., {Hollenbach}, D.~J., {Lord}, S.~D., {et~al.} 1994, \apj, 423, 223

\bibitem[{{Carter} {et~al.}(2012){Carter}, {Lazareff}, {Maier}, {Chenu},
  {Fontana}, {Bortolotti}, {Boucher}, {Navarrini}, {Blanchet}, {Greve}, {John},
  {Kramer}, {Morel}, {Navarro}, {Pe{\~n}alver}, {Schuster}, \&
  {Thum}}]{carter12}
{Carter}, M., {Lazareff}, B., {Maier}, D., {et~al.} 2012, \aap, 538, A89

\bibitem[{{Castor} {et~al.}(1975){Castor}, {McCray}, \& {Weaver}}]{castor75}
{Castor}, J., {McCray}, R., \& {Weaver}, R. 1975, \apjl, 200, L107

\bibitem[{{Chini} {et~al.}(1980){Chini}, {Els{\"a}sser}, \& {Neckel}}]{chini80}
{Chini}, R., {Els{\"a}sser}, H., \& {Neckel}, T. 1980, \aap, 91, 186

\bibitem[{{Clark} {et~al.}(2011){Clark}, {Glover}, {Klessen}, \&
  {Bromm}}]{clark11}
{Clark}, P.~C., {Glover}, S.~C.~O., {Klessen}, R.~S., \& {Bromm}, V. 2011,
  \apj, 727, 110

\bibitem[{{Dale} \& {Bonnell}(2011)}]{dale11}
{Dale}, J.~E. \& {Bonnell}, I. 2011, \mnras, 414, 321

\bibitem[{{Federrath} \& {Klessen}(2012)}]{federrath12}
{Federrath}, C. \& {Klessen}, R.~S. 2012, \apj, 761, 156

\bibitem[{{Frerking} {et~al.}(1989){Frerking}, {Keene}, {Blake}, \&
  {Phillips}}]{frerking89}
{Frerking}, M.~A., {Keene}, J., {Blake}, G.~A., \& {Phillips}, T.~G. 1989,
  \apj, 344, 311

\bibitem[{{Genzel} {et~al.}(1988){Genzel}, {Harris}, {Stutzki}, \&
  {Jaffe}}]{genzel88}
{Genzel}, R., {Harris}, A.~I., {Stutzki}, J., \& {Jaffe}, D.~T. 1988, \apj,
  332, 1049

\bibitem[{{Gerin} \& {Phillips}(1998)}]{gerin98}
{Gerin}, M. \& {Phillips}, T.~G. 1998, \apjl, 509, L17

\bibitem[{{Goodwin}(1997)}]{goodwin97}
{Goodwin}, S.~P. 1997, \mnras, 284, 785

\bibitem[{{Gordon}(1988)}]{gordon88}
{Gordon}, M.~A. 1988, {H II regions and radio recombination lines}, ed. K.~I.
  {Kellermann} \& G.~L. {Verschuur}, 37--94

\bibitem[{{Gordon}(1989)}]{gordon89}
{Gordon}, M.~A. 1989, \apj, 337, 782

\bibitem[{{Gordon} \& {Sorochenko}(2002)}]{gordon02}
{Gordon}, M.~A. \& {Sorochenko}, R.~L., eds. 2002, Astrophysics and Space
  Science Library, Vol. 282, {Radio Recombination Lines. Their Physics and
  Astronomical Applications}

\bibitem[{{Graf} {et~al.}(1993){Graf}, {Eckart}, {Genzel}, {Harris},
  {Poglitsch}, {Russell}, \& {Stutzki}}]{graf93}
{Graf}, U.~U., {Eckart}, A., {Genzel}, R., {et~al.} 1993, \apj, 405, 249

\bibitem[{{Griem}(1967)}]{griem67}
{Griem}, H.~R. 1967, \apj, 148, 547

\bibitem[{{Griffin} {et~al.}(1986){Griffin}, {Ade}, {Orton}, {Robson}, {Gear},
  {Nolt}, \& {Radostitz}}]{griffin86}
{Griffin}, M.~J., {Ade}, P.~A.~R., {Orton}, G.~S., {et~al.} 1986, Icarus, 65,
  244

\bibitem[{{G\"usten} \& {Fiebig}(1988)}]{gusten88}
{G\"usten}, R. \& {Fiebig}, D. 1988, \aap, 204, 253

\bibitem[{{G{\"u}sten} {et~al.}(2006){G{\"u}sten}, {Nyman}, {Schilke},
  {Menten}, {Cesarsky}, \& {Booth}}]{gusten06}
{G{\"u}sten}, R., {Nyman}, L.~{\AA}., {Schilke}, P., {et~al.} 2006, \aap, 454,
  L13

\bibitem[{{Habing}(1968)}]{habing68}
{Habing}, H.~J. 1968, \bain, 19, 421

\bibitem[{{Hanson} {et~al.}(1997){Hanson}, {Howarth}, \& {Conti}}]{hanson97}
{Hanson}, M.~M., {Howarth}, I.~D., \& {Conti}, P.~S. 1997, \apj, 489, 698

\bibitem[{{Harris} {et~al.}(1987){Harris}, {Stutzki}, {Genzel}, {Lugten},
  {Stacey}, \& {Jaffe}}]{harris87}
{Harris}, A.~I., {Stutzki}, J., {Genzel}, R., {et~al.} 1987, \apjl, 322, L49

\bibitem[{{Henley} {et~al.}(2012){Henley}, {Kwak}, \& {Shelton}}]{henley12}
{Henley}, D.~B., {Kwak}, K., \& {Shelton}, R.~L. 2012, \apj, 753, 58

\bibitem[{{Heyminck} {et~al.}(2012){Heyminck}, {Graf}, {G{\"u}sten}, {Stutzki},
  {H{\"u}bers}, \& {Hartogh}}]{heyminck12}
{Heyminck}, S., {Graf}, U.~U., {G{\"u}sten}, R., {et~al.} 2012, \aap, 542, L1

\bibitem[{{Heyminck} {et~al.}(2006){Heyminck}, {Kasemann}, {G{\"u}sten}, {de
  Lange}, \& {Graf}}]{heyminck06}
{Heyminck}, S., {Kasemann}, C., {G{\"u}sten}, R., {de Lange}, G., \& {Graf},
  U.~U. 2006, \aap, 454, L21

\bibitem[{{Hobson}(1992)}]{hobson92}
{Hobson}, M.~P. 1992, \mnras, 256, 457

\bibitem[{{Hocuk} \& {Spaans}(2010)}]{hocuk10}
{Hocuk}, S. \& {Spaans}, M. 2010, \aap, 522, A24+

\bibitem[{{Hoffmeister} {et~al.}(2008){Hoffmeister}, {Chini}, {Scheyda},
  {Schulze}, {Watermann}, {N{\"u}rnberger}, \& {Vogt}}]{hoffmeister08}
{Hoffmeister}, V.~H., {Chini}, R., {Scheyda}, C.~M., {et~al.} 2008, \apj, 686,
  310

\bibitem[{{Hollenbach} {et~al.}(1991){Hollenbach}, {Takahashi}, \&
  {Tielens}}]{hollenbach91}
{Hollenbach}, D.~J., {Takahashi}, T., \& {Tielens}, A.~G.~G.~M. 1991, \apj,
  377, 192

\bibitem[{{Hollenbach} \& {Tielens}(1999)}]{hollenbach99}
{Hollenbach}, D.~J. \& {Tielens}, A.~G.~G.~M. 1999, Reviews of Modern Physics,
  71, 173

\bibitem[{{Howe} {et~al.}(2000){Howe}, {Ashby}, {Bergin}, {Chin}, {Erickson},
  {Goldsmith}, \& {et al.}}]{howe00}
{Howe}, J.~E., {Ashby}, M.~L.~N., {Bergin}, E.~A., {et~al.} 2000, \apjl, 539,
  L137

\bibitem[{{Jaffe} {et~al.}(1987){Jaffe}, {Harris}, \& {Genzel}}]{jaffe87}
{Jaffe}, D.~T., {Harris}, A.~I., \& {Genzel}, R. 1987, \apj, 316, 231

\bibitem[{{Johnson} {et~al.}(1998){Johnson}, {Depree}, \& {Goss}}]{johnson98}
{Johnson}, C.~O., {Depree}, C.~G., \& {Goss}, W.~M. 1998, \apj, 500, 302

\bibitem[{{Keene} {et~al.}(1985){Keene}, {Blake}, {Phillips}, {Huggins}, \&
  {Beichman}}]{keene85}
{Keene}, J., {Blake}, G.~A., {Phillips}, T.~G., {Huggins}, P.~J., \&
  {Beichman}, C.~A. 1985, \apj, 299, 967

\bibitem[{{Klein} {et~al.}(2012){Klein}, {Hochg{\"u}rtel}, {Kr{\"a}mer},
  {Bell}, {Meyer}, \& {G{\"u}sten}}]{klein12}
{Klein}, B., {Hochg{\"u}rtel}, S., {Kr{\"a}mer}, I., {et~al.} 2012, \aap, 542,
  L3

\bibitem[{{Kleinmann}(1973)}]{kleinmann73}
{Kleinmann}, D.~E. 1973, \aplett, 13, 49

\bibitem[{{Klessen} {et~al.}(2005){Klessen}, {Ballesteros-Paredes},
  {V{\'a}zquez-Semadeni}, \& {Dur{\'a}n-Rojas}}]{klessen05}
{Klessen}, R.~S., {Ballesteros-Paredes}, J., {V{\'a}zquez-Semadeni}, E., \&
  {Dur{\'a}n-Rojas}, C. 2005, \apj, 620, 786

\bibitem[{{Kramer} {et~al.}(2008){Kramer}, {Cubick}, {R{\"o}llig}, {Sun},
  {Yonekura}, \& {et al.}}]{kramer08}
{Kramer}, C., {Cubick}, M., {R{\"o}llig}, M., {et~al.} 2008, \aap, 477, 547

\bibitem[{{Kramer} {et~al.}(2004){Kramer}, {Jakob}, {Mookerjea}, {Schneider},
  {Br{\"u}ll}, \& {Stutzki}}]{kramer04}
{Kramer}, C., {Jakob}, H., {Mookerjea}, B., {et~al.} 2004, \aap, 424, 887

\bibitem[{{Kramer} {et~al.}(1998){Kramer}, {Stutzki}, {Rohrig}, \&
  {Corneliussen}}]{kramer98}
{Kramer}, C., {Stutzki}, J., {Rohrig}, R., \& {Corneliussen}, U. 1998, \aap,
  329, 249

\bibitem[{{Krumholz} {et~al.}(2010){Krumholz}, {Cunningham}, {Klein}, \&
  {McKee}}]{krumholz10}
{Krumholz}, M.~R., {Cunningham}, A.~J., {Klein}, R.~I., \& {McKee}, C.~F. 2010,
  \apj, 713, 1120

\bibitem[{{Lada} {et~al.}(1991){Lada}, {Depoy}, {Merrill}, \&
  {Gatley}}]{lada91}
{Lada}, C.~J., {Depoy}, D.~L., {Merrill}, K.~M., \& {Gatley}, I. 1991, \apj,
  374, 533

\bibitem[{{Loren}(1989)}]{loren89}
{Loren}, R.~B. 1989, \apj, 338, 902

\bibitem[{{Luhman} {et~al.}(2003){Luhman}, {Satyapal}, {Fischer}, {Wolfire},
  {Sturm}, {Dudley}, {Lutz}, \& {Genzel}}]{luhman03}
{Luhman}, M.~L., {Satyapal}, S., {Fischer}, J., {et~al.} 2003, \apj, 594, 758

\bibitem[{{Meijerink} \& {Spaans}(2005)}]{meijerink05}
{Meijerink}, R. \& {Spaans}, M. 2005, \aap, 436, 397

\bibitem[{{Meijerink} {et~al.}(2007){Meijerink}, {Spaans}, \&
  {Israel}}]{meijerink07}
{Meijerink}, R., {Spaans}, M., \& {Israel}, F.~P. 2007, \aap, 461, 793

\bibitem[{{Meixner} {et~al.}(1992){Meixner}, {Haas}, {Tielens}, {Erickson}, \&
  {Werner}}]{meixner92}
{Meixner}, M., {Haas}, M., {Tielens}, A., {Erickson}, E., \& {Werner}, M. 1992,
  \apj, 390, 499

\bibitem[{{Mookerjea} {et~al.}(2003){Mookerjea}, {Ghosh}, {Kaneda}, {Nakagawa},
  {Ojha}, {Rengarajan}, {Shibai}, \& {Verma}}]{mookerjea03}
{Mookerjea}, B., {Ghosh}, S.~K., {Kaneda}, H., {et~al.} 2003, \aap, 404, 569

\bibitem[{{Mookerjea} {et~al.}(2011){Mookerjea}, {Kramer}, {Buchbender},
  {Boquien}, {Verley}, {Rela{\~n}o}, {Quintana-Lacaci}, {Aalto}, {Braine},
  {Calzetti}, {Combes}, {Garcia-Burillo}, {Gratier}, {Henkel}, {Israel},
  {Lord}, {Nikola}, {R{\"o}llig}, {Stacey}, {Tabatabaei}, {van der Tak}, \&
  {van der Werf}}]{mookerjea11}
{Mookerjea}, B., {Kramer}, C., {Buchbender}, C., {et~al.} 2011, \aap, 532, A152

\bibitem[{{Muders} {et~al.}(2006){Muders}, {Hafok}, {Wyrowski}, {Polehampton},
  {Belloche}, {K{\"o}nig}, {Schaaf}, {Schuller}, {Hatchell}, \& {van der
  Tak}}]{muders06}
{Muders}, D., {Hafok}, H., {Wyrowski}, F., {et~al.} 2006, \aap, 454, L25

\bibitem[{{Neufeld} {et~al.}(2010){Neufeld}, {Sonnentrucker}, {Phillips},
  {Lis}, {de Luca}, {Goicoechea}, {Black}, {Gerin}, {Bell}, {Boulanger},
  {Cernicharo}, {Coutens}, {Dartois}, {Kazmierczak}, {Encrenaz}, {Falgarone},
  {Geballe}, {Giesen}, {Godard}, {Goldsmith}, {Gry}, {Gupta}, {Hennebelle},
  {Herbst}, {Hily-Blant}, {Joblin}, {Ko{\l}os}, {Kre{\l}owski},
  {Mart{\'{\i}}n-Pintado}, {Menten}, {Monje}, {Mookerjea}, {Pearson},
  {Perault}, {Persson}, {Plume}, {Salez}, {Schlemmer}, {Schmidt}, {Stutzki},
  {Teyssier}, {Vastel}, {Yu}, {Cais}, {Caux}, {Liseau}, {Morris}, \&
  {Planesas}}]{neufeld10}
{Neufeld}, D.~A., {Sonnentrucker}, P., {Phillips}, T.~G., {et~al.} 2010, \aap,
  518, L108

\bibitem[{{Pearson}(1920)}]{pearson20}
{Pearson}, K. 1920, Biometrica, 13, 25

\bibitem[{{Peimbert} {et~al.}(1988){Peimbert}, {Ukita}, {Hasegawa}, \&
  {Jugaku}}]{peimbert88}
{Peimbert}, M., {Ukita}, N., {Hasegawa}, T., \& {Jugaku}, J. 1988, \pasj, 40,
  581

\bibitem[{{Pellegrini} {et~al.}(2007){Pellegrini}, {Baldwin}, {Brogan},
  {Hanson}, {Abel}, {Ferland}, {Nemala}, {Shaw}, \& {Troland}}]{pellegrini07}
{Pellegrini}, E.~W., {Baldwin}, J.~A., {Brogan}, C.~L., {et~al.} 2007, \apj,
  658, 1119

\bibitem[{{P{\'e}rez-Beaupuits} {et~al.}(2007){P{\'e}rez-Beaupuits}, {Aalto},
  \& {Gerebro}}]{pb07}
{P{\'e}rez-Beaupuits}, J.~P., {Aalto}, S., \& {Gerebro}, H. 2007, \aap, 476,
  177

\bibitem[{{P{\'e}rez-Beaupuits} {et~al.}(2010){P{\'e}rez-Beaupuits}, {Spaans},
  {Hogerheijde}, {G{\"u}sten}, {Baryshev}, \& {Boland}}]{pb10}
{P{\'e}rez-Beaupuits}, J.~P., {Spaans}, M., {Hogerheijde}, M.~R., {et~al.}
  2010, \aap, 510, A87+

\bibitem[{{P{\'e}rez-Beaupuits} {et~al.}(2009){P{\'e}rez-Beaupuits}, {Spaans},
  {van der Tak}, {Aalto}, {Garc{\'{\i}}a-Burillo}, {Fuente}, \& {Usero}}]{pb09}
{P{\'e}rez-Beaupuits}, J.~P., {Spaans}, M., {van der Tak}, F.~F.~S., {et~al.}
  2009, \aap, 503, 459

\bibitem[{{P{\'e}rez-Beaupuits} {et~al.}(2013){P{\'e}rez-Beaupuits}, {Stutzki},
  {G{\"u}sten}, {Ossenkopf}, \& {Wiesemeyer}}]{pb13}
{P{\'e}rez-Beaupuits}, J.~P., {Stutzki}, J., {G{\"u}sten}, R., {Ossenkopf}, V.,
  \& {Wiesemeyer}, H. 2013, in IAU Symposium, Vol. 292, IAU Symposium, ed.
  T.~{Wong} \& J.~{Ott}, 55--55

\bibitem[{{P{\'e}rez-Beaupuits} {et~al.}(2012){P{\'e}rez-Beaupuits},
  {Wiesemeyer}, {Ossenkopf}, {Stutzki}, {G{\"u}sten}, {Simon}, {H{\"u}bers}, \&
  {Ricken}}]{pb12}
{P{\'e}rez-Beaupuits}, J.~P., {Wiesemeyer}, H., {Ossenkopf}, V., {et~al.} 2012,
  \aap, 542, L13

\bibitem[{{P{\'e}rez-Beaupuits} {et~al.}(, in prep.){P{\'e}rez-Beaupuits},
  {Zinnecker}, {G{\"u}sten}, {Requena-Torres}, {H{\"u}bers}, \&
  {Ricken}}]{pb15}
{P{\'e}rez-Beaupuits}, J.~P., {Zinnecker}, H., {G{\"u}sten}, R., {et~al.} , in
  prep.

\bibitem[{{Poelman} \& {Spaans}(2005)}]{poelman05}
{Poelman}, D.~R. \& {Spaans}, M. 2005, \aap, 440, 559

\bibitem[{{Rodgers} \& {Nicewander}(1988)}]{rodgers88}
{Rodgers}, J.~L. \& {Nicewander}, W.~A. 1988, The American Statistician, 42, 59

\bibitem[{{R{\"o}llig} {et~al.}(2011){R{\"o}llig}, {Kramer}, {Rajbahak},
  {Minamidani}, {Sun}, {Simon}, {Ossenkopf}, {Cubick}, {Hitschfeld}, {Aravena},
  {Bensch}, {Bertoldi}, {Bronfman}, {Fujishita}, {Fukui}, {Graf}, {Honingh},
  {Ito}, {Jakob}, {Jacobs}, {Klein}, {Koo}, {May}, {Miller}, {Miyamoto},
  {Mizuno}, {Onishi}, {Park}, {Pineda}, {Rabanus}, {Sasago}, {Schieder},
  {Stutzki}, {Yamamoto}, \& {Yonekura}}]{roellig11}
{R{\"o}llig}, M., {Kramer}, C., {Rajbahak}, C., {et~al.} 2011, \aap, 525, A8

\bibitem[{{Russell} {et~al.}(1981){Russell}, {Melnick}, {Smyers}, {Kurtz},
  {Gosnell}, {Harwit}, \& {Werner}}]{russell81}
{Russell}, R.~W., {Melnick}, G., {Smyers}, S.~D., {et~al.} 1981, \apjl, 250,
  L35

\bibitem[{{Sargsyan} {et~al.}(2014){Sargsyan}, {Samsonyan}, {Lebouteiller},
  {Weedman}, {Barry}, {Bernard-Salas}, {Houck}, \& {Spoon}}]{sargsyan14}
{Sargsyan}, L., {Samsonyan}, A., {Lebouteiller}, V., {et~al.} 2014, \apj, 790,
  15

\bibitem[{{Schneider} {et~al.}(2003){Schneider}, {Simon}, {Kramer}, {Kraemer},
  {Stutzki}, \& {Mookerjea}}]{schneider03}
{Schneider}, N., {Simon}, R., {Kramer}, C., {et~al.} 2003, \aap, 406, 915

\bibitem[{{Schneider} {et~al.}(2002){Schneider}, {Simon}, {Kramer}, {Stutzki},
  \& {Bontemps}}]{schneider02}
{Schneider}, N., {Simon}, R., {Kramer}, C., {Stutzki}, J., \& {Bontemps}, S.
  2002, \aap, 384, 225

\bibitem[{{Sch{\"o}ier} {et~al.}(2005){Sch{\"o}ier}, {van der Tak}, {van
  Dishoeck}, \& {Black}}]{schoier05}
{Sch{\"o}ier}, F.~L., {van der Tak}, F.~F.~S., {van Dishoeck}, E.~F., \&
  {Black}, J.~H. 2005, \aap, 432, 369

\bibitem[{{Spaans}(1996)}]{spaans96}
{Spaans}, M. 1996, \aap, 307, 271

\bibitem[{{Spaans} \& {van Dishoeck}(1997)}]{spaans97}
{Spaans}, M. \& {van Dishoeck}, E.~F. 1997, \aap, 323, 953

\bibitem[{{Stacey} {et~al.}(1991){Stacey}, {Geis}, {Genzel}, {Lugten},
  {Poglitsch}, {Sternberg}, \& {Townes}}]{stacey91}
{Stacey}, G.~J., {Geis}, N., {Genzel}, R., {et~al.} 1991, \apj, 373, 423

\bibitem[{{Stutzki} \& {G\"usten}(1990)}]{stutzki90}
{Stutzki}, J. \& {G\"usten}, R. 1990, \apj, 356, 513

\bibitem[{{Stutzki} {et~al.}(1988){Stutzki}, {Stacey}, {Genzel}, {Harris},
  {Jaffe}, \& {Lugten}}]{stutzki88}
{Stutzki}, J., {Stacey}, G.~J., {Genzel}, R., {et~al.} 1988, \apj, 332, 379

\bibitem[{{Tenorio-Tagle}(1979)}]{tenorio79}
{Tenorio-Tagle}, G. 1979, \aap, 71, 59

\bibitem[{{Tsivilev} \& {Krasnov}(1999)}]{tsivilev99}
{Tsivilev}, A.~P. \& {Krasnov}, V.~V. 1999, Astronomy Reports, 43, 511

\bibitem[{{van der Tak} {et~al.}(2007){van der Tak}, {Black}, {Sch{\"o}ier},
  {Jansen}, \& {van Dishoeck}}]{vdtak07}
{van der Tak}, F.~F.~S., {Black}, J.~H., {Sch{\"o}ier}, F.~L., {Jansen}, D.~J.,
  \& {van Dishoeck}, E.~F. 2007, \aap, 468, 627

\bibitem[{{Wakelam} \& {Herbst}(2008)}]{wakelam08}
{Wakelam}, V. \& {Herbst}, E. 2008, \apj, 680, 371

\bibitem[{{Weaver} {et~al.}(1977){Weaver}, {McCray}, {Castor}, {Shapiro}, \&
  {Moore}}]{weaver77}
{Weaver}, R., {McCray}, R., {Castor}, J., {Shapiro}, P., \& {Moore}, R. 1977,
  \apj, 218, 377

\bibitem[{{Wolfire} {et~al.}(2010){Wolfire}, {Hollenbach}, \&
  {McKee}}]{wolfire10}
{Wolfire}, M.~G., {Hollenbach}, D., \& {McKee}, C.~F. 2010, \apj, 716, 1191

\bibitem[{{Yamamoto} {et~al.}(2001){Yamamoto}, {Maezawa}, {Ikeda}, {Ito},
  {Oka}, \& {et al.}}]{yamamoto01}
{Yamamoto}, S., {Maezawa}, H., {Ikeda}, M., {et~al.} 2001, \apjl, 547, L165

\end{thebibliography}


\Online

\normalsize

\begin{appendix}

\section{LTE Analysis of \ci}\label{sec:appendix-0}

\normalfont

\begin{figure}[!tp]

  \hfill\includegraphics[,angle=0,width=0.5\textwidth]{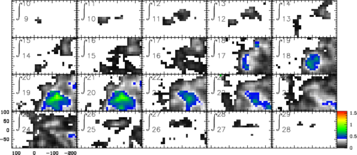}\hspace*{\fill}\\

  \caption{\footnotesize{Velocity channel maps at 1~\kms\ width of the optical depth $\tau_{1\to0}$ of the \ci~$1\to0$ line, estimated assuming LTE conditions in M17~SW.}
}

  \label{fig:CI-Tau10}
\end{figure}

\begin{figure}[!tp]

  \hfill\includegraphics[,angle=0,width=0.5\textwidth]{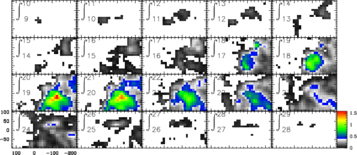}\hspace*{\fill}\\

  \caption{\footnotesize{Velocity channel maps at 1~\kms\ width of the optical depth $\tau_{2\to1}$ of the \ci~$2\to1$ line, estimated assuming LTE conditions in M17~SW.}
}

  \label{fig:CI-Tau21}
\end{figure}

\begin{figure}[!tp]

  \hfill\includegraphics[,angle=0,width=0.5\textwidth]{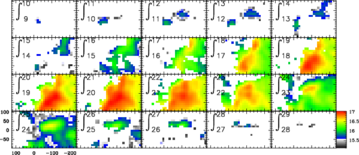}\hspace*{\fill}\\

  \caption{\footnotesize{Velocity channel maps at 1~\kms\ width of the column density ($\2cm$) of \ci\ (in $log_{10}$ scale), estimated using the excitation temperature from Fig.~\ref{fig:CI-Tex}}.
}

  \label{fig:CI-column}
\end{figure}

Following \citet{frerking89} and Schneider \etal\ (2003, their Appendix A), the optical depths of the \ci\ 
$J=1\rightarrow0$ and $J=2\rightarrow1$ lines can be estimated from the excitation temperature and the peak 
intensity of both lines, assuming a beam filling factor of unity. Knowing the excitation temperature and the 
optical depth of the $J=1\rightarrow0$ line, the column density $N($\ci$)$ can be computed as well. 
Figures~\ref{fig:CI-Tau10} and \ref{fig:CI-Tau21} shows the channel maps corresponding to the optical depths of 
both \ci\ lines, while Fig.~\ref{fig:CI-column} shows the channel maps of the \ci\ column density, as estimated 
assuming LTE conditions.

\section{Non-LTE Excitation Analysis of \ci}\label{sec:appendix-A}

Following the procedure described in \citep{pb07, pb09}, we use the radiative transfer code 
RADEX \citep{vdtak07} to create a data cube containing the intensities, as well as the excitation 
temperatures and optical depths of the two \ci\ transitions, in function of a range of kinetic 
temperatures $T_K$, number densities $n({\rm H_2})$ (i.e., excitation conditions), and column densities per line 
width $N/\Delta V$. The collision rates used were taken from the LAMDA database \citep{schoier05}.

We only used  collisions with H$_2$  since this is the most abundant molecule. Other 
collision partners can be H and He. Although their collision cross sections are comparable, H$_2$ is about 5 
times more abundant than He, and H is at least one order of magnitude less abundant than H$_2$ in the dense cores 
of molecular clouds \citep[e.g.,][]{meijerink05}. Hence, including H and He as additional collision partners 
would not produce a significant change in our results. We also assumed a homogeneous spherical symmetry in the 
clumps for the escape-probability formalism.

The original RADEX code was modified  to include dust background emission as a diluted blackbody 
radiation field, as in \citet{poelman05} and \citet{pb09}.
The total background radiation is modeled as a composite between the cosmic background radiation (CMB), assumed 
to be a blackbody function at 2.73 K, and the diluted dust radiation estimated as 
$\tau_{dust}\times B(T_{dust})$, where $B(T_{dust})$ is the Planck function and the dust continuum optical depth 
$\tau_{dust}(\lambda)$ is defined by \citet{hollenbach91} as 
$\tau_{dust}(\lambda)=\tau_{100\mu{\rm m}}(100\mu{\rm m}/\lambda)$. 
We adopted an average dust background temperature $T_{dust}=50$ K and the high FIR opacity 
$\tau_{100\mu {\rm m}}=0.106$ found by \citet{meixner92} in M17SW.

\begin{figure}[!tp]
\centering

\epsfig{file=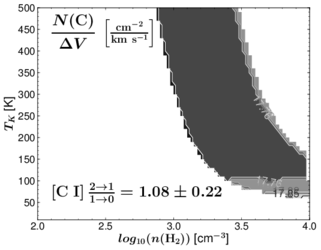,width=0.9\linewidth,clip=}

\begin{tabular}{ll}

\hspace{-0.3cm}\epsfig{file=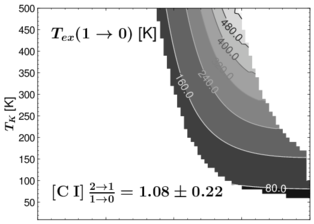,width=0.53\linewidth,clip=} &
\hspace{-0.45cm}\epsfig{file=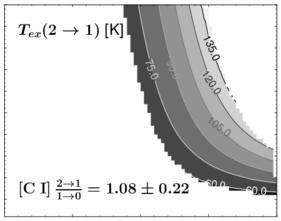,width=0.475\linewidth,clip=} \\[-0.15cm]

\hspace{-0.3cm}\epsfig{file=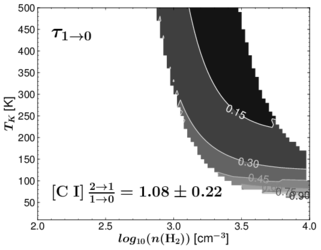,width=0.54\linewidth,clip=} &
\hspace{-0.55cm}\epsfig{file=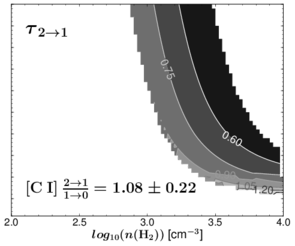,width=0.497\linewidth,clip=} \\[-0.15cm]

\end{tabular}

\caption{\footnotesize{Excitation map (\textit{top panel}) for the \ci~$\frac{2\to1}{1\to0}$ line ratio observed at offset position ($-130''$,$-10''$) in the velocity channel 19--20~\kms. The contours and labels correspond to the column density per line width $N/\Delta V$ (\ndv, in $log_{10}$ scale). In the \textit{middle panel} the contours and labels correspond to the excitation temperature $T_{ex}$ (K) of the $J=1\to0$ (\textit{left}) and $J=2\to1$ (\textit{right}) transitions, while the {\it bottom panel} shows the corresponding optical depths for each line.}}

\label{fig:CI-xcmaps-pos1}
\end{figure}

\begin{figure}[!tp]
\centering

\epsfig{file=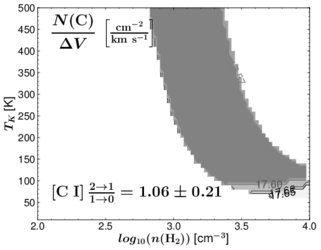,width=0.9\linewidth,clip=}

\begin{tabular}{ll}

\hspace{-0.3cm}\epsfig{file=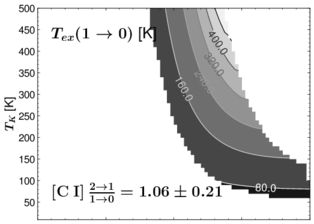,width=0.53\linewidth,clip=} &
\hspace{-0.45cm}\epsfig{file=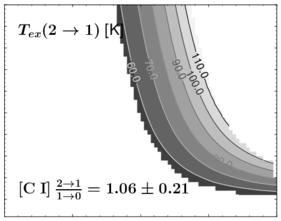,width=0.475\linewidth,clip=} \\[-0.15cm]

\hspace{-0.3cm}\epsfig{file=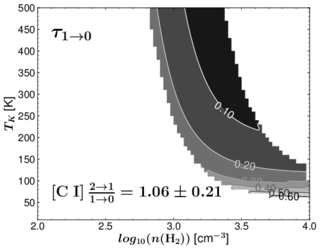,width=0.54\linewidth,clip=} &
\hspace{-0.55cm}\epsfig{file=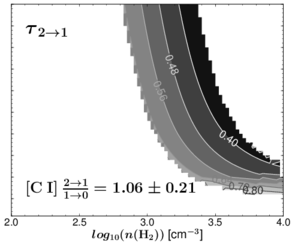,width=0.497\linewidth,clip=} \\[-0.15cm]
\end{tabular}

\caption{\footnotesize{Excitation map (\textit{top panel}) for the \ci~$\frac{2\to1}{1\to0}$ line ratio observed at offset position ($-130''$,$-70''$) in the velocity channel 19--20~\kms. The contours and labels correspond to the column density per line width $N/\Delta V$ (\ndv, in $log_{10}$ scale). In the \textit{middle panel} the contours and labels correspond to the excitation temperature $T_{ex}$ (K) of the $J=1\to0$ (\textit{left}) and $J=2\to1$ (\textit{right}) transitions, while the {\it bottom panel} shows the corresponding optical depths for each line.}}

\label{fig:CI-xcmaps-pos2}
\end{figure}

We assume that the emission collected by the beam has a homogeneous elliptical Gaussian distribution and 
that the coupling factor of our beam to the source distribution is unity, so we can compare directly with the 
output of RADEX, which is the Rayleigh-Jeans equivalent radiation temperature $T_R$ emitted by the source.

We explored all the possible excitation conditions within the given range that can lead to the observed  
radiation temperatures and the line ratios between the two \ci\ lines.
 The line ratios and the peak temperature of the lower-$J$ line involved in each ratio were used 
to constrain the excitation conditions.   
Including the rms of the observed spectra and uncertainties in all the assumptions mentioned 
above, a 20\% error of the ratios and peak temperatures are used to define a range of values for 
$T_K$, $n({\rm H_2})$, and $N/\Delta V$ within which the RADEX output is selected as a valid solution.

The volume density explored ranges between $10^2~\3cm$ and $10^4~\3cm$, the kinetic temperature varies from 10 K 
to 500 K, and the column density per line width lies between $10^{10}$ \ndv\ and $10^{20}$ \ndv. Since the 
optical depth, and the line intensities, are proportional to the column density per line width $N/\Delta V$, we 
generated the RADEX data cube assuming the $\Delta V=1$~\kms\ of the velocity channels in the re-sampled 
spectra. In order to constrain the solutions, we fit the line ratio between the peak temperatures of the 
transitions and the radiation temperature of the lower transition ($J=1\to0$), which are $R\sim1.08$ and 
$T_R\approx T_{mb}=31.9$~K at offset position ($-130''$,$-10''$), and $R\sim1.06$ and $T_R\approx T_{mb}=22.4$~K 
at offset position ($-130''$,$-70''$).

Figures~\ref{fig:CI-xcmaps-pos1} to \ref{fig:CI-xcmaps-pos2} show gray scale and contour {\it \emph{maps}} of 
the excitation conditions found to reproduce the observed \ci\ line ratios and peak temperatures at the offset 
positions ($-130''$,$-10''$) and ($-130''$,$-70''$), respectively. The values shown correspond to the average of 
all the possible $N/\Delta V$, $T_{ex}$, and $\tau$ found for each pair of excitation conditions ($n(\rm H_2)$ 
and $T_K$).
The curvature in the solutions depict the dichotomy between the kinetic temperature and the density of 
the collision partner. In other words, solutions for the observed values can be found for higher 
temperatures and lower densities, but also for lower $T_K$ and higher $n({\rm H_2})$. The column density per 
line width does not change significantly along the solution curve, but it does change slightly across the 
curves, and especially at lower ($<100$~K) kinetic temperatures. This means that for a given kinetic temperature, 
$N/\Delta V$ will show a small variation in function of density.

\section{Correlation between line tracers}\label{sec:appendix-B}

\begin{figure*}[!ht]

  \hfill\includegraphics[angle=0,width=0.24\textwidth]{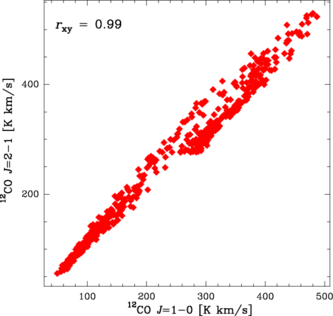}%
  \hfill\includegraphics[angle=0,width=0.24\textwidth]{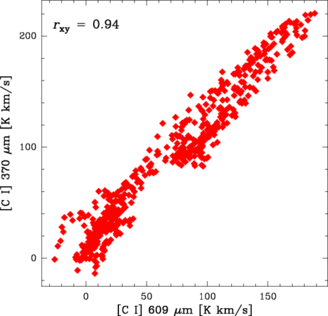}%
  \hfill\includegraphics[angle=0,width=0.24\textwidth]{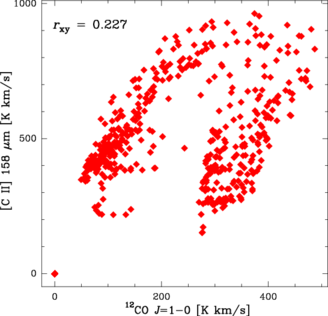}\hspace*{\fill}\\

  \caption{\footnotesize{Example of the scatter plots between the pixel values of the velocity-integrated 
  intensity maps of two different line tracers, and the corresponding correlation coefficient obtained using 
  Eq.~\ref{eq:corr}.}}

  \label{fig:scatter-area}
\end{figure*}

\begin{figure*}[!ht]

  \hfill\includegraphics[angle=0,width=0.48\textwidth]{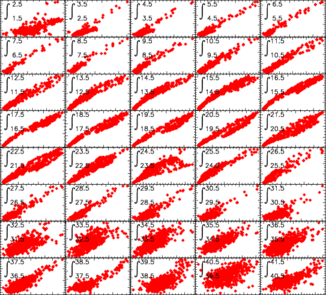}%
  \hfill\includegraphics[angle=0,width=0.48\textwidth]{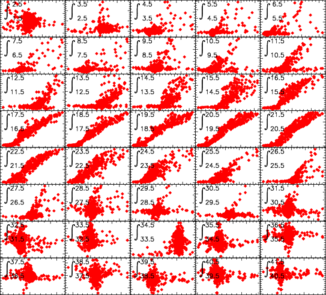}\hspace*{\fill}\\  

  \hfill\includegraphics[angle=0,width=0.48\textwidth]{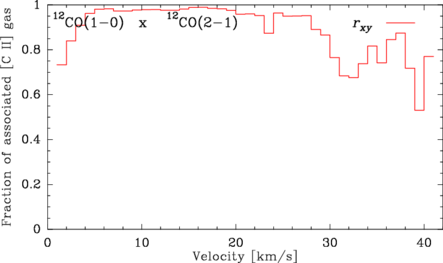}%
  \hfill\includegraphics[angle=0,width=0.48\textwidth]{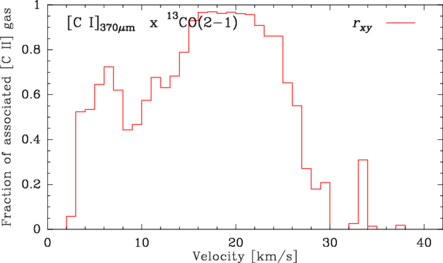}\hspace*{\fill}\\

  \caption{\footnotesize{Example of the scatter plots between the pixel values (\Kkms) of the velocity channel maps of the \twco\  $J=1\to0$ (X-axis) and $J=2-1$ (Y-axis) transitions (\textit{top left}), and the \ci~370~\mum\ (X-axis) and \thco~$J=2\to1$ (Y-axis) lines (\textit{top right}). The corresponding correlation coefficient $r$ at each velocity channel, is shown in the bottom panels.}}

  \label{fig:scatter-chan}
\end{figure*}

In order to verify the \textit{\emph{apparent}} spatial correlation between two species, we use the correlation between 
a specific pixel value of a map, and the same pixel in the map of another tracer. This can be done only in the 
maps with the same dimensions and spatial resolutions. Therefore, we first convolved all the maps to the larger 
beam size (24$''$) of the \ceio~$J=1-0$ line, and we used the SOFIA/GREAT map of the \cii~158~\mum\ line (which 
covers the smaller region) as a template to create all the \ci\ and CO maps. 
The sample correlation coefficient commonly used to estimate $r_{xy}$ between the images $X$ and $Y$ is the 
Pearson's product-moment correlation coefficient defined by \citep{pearson20, rodgers88},

\begin{equation}\label{eq:corr}
r_{xy} = \frac{cov[X,Y]}{\sqrt{var[X]var[Y]}} = \frac{\sum_i \sum_j [X_{ij}-\bar{X}] [Y_{ij}-\bar{Y}]}{\sqrt{\sum_i \sum_j [X_{ij}-\bar{X}]^2 \sum_i \sum_j [Y_{ij}-\bar{Y}]^2}}
,\end{equation}

\noindent
where $X_{ij}$ and $Y_{ij}$ are the pixel values of two given maps or images (e.g., \cii\ and \twco~$J=1\to0$), 
and $\bar{X}$ and $\bar{Y}$ are the average pixel value of the respective maps. We note that the uncertainty of the 
correlation coefficient can be approximated as $\sigma_{r}\sim(1-r_{xy}^2)/\sqrt{n-2}$, and since we have maps 
with $n=31\times18=558$ pixels, the uncertainty will always be between 10$^{-2}$ and 10$^{-3}$. 
This correlation is applied to the velocity integrated intensity maps,  and to the 1~\kms\ width channel maps. 
We note that because of the positive correlation between the intensities of different tracers, $r_{xy}$ ranges 
between 0 and 1.
As proof of concept, the scatter plot and the associated correlation coefficient between the 
velocity-integrated intensity of several line tracers is shown in Fig.~\ref{fig:scatter-area}. The application 
of the scatter plot to the channel maps is shown in Fig.~\ref{fig:scatter-chan} for two test cases. 
We note that the correlation between the velocity-integrated intensity, as well as the channel maps, of the 
$J=1\to0$ and $J=2\to1$ transitions of \twco\ is very good, even at the faintest emission of the lower velocity 
($>3.5~\kms$) channels,  but the good correlation is lost at the higher velocity ($>28.5~\kms$) channels. In the 
case of the \ci~370~\mum\ (X-axis) versus \thco~$J=2\to1$ (Y-axis), the pixel values are more scattered at the 
lower ($<16$~\kms) and higher ($>22$~\kms) velocity channels because the line intensities are fainter and, hence, 
those channel maps are more affected by noise.

\section{Spatial association in channel maps}\label{sec:appendix-C}

We use a simple method to estimate the fraction of the region mapped where two emission lines are associated in 
narrow (1~\kms) width channel maps. All maps are first convolved to the lowest spatial resolution (24$''$ FWHM 
beam) of the \ceio~$J=1\to0$ map in order to increase the S/N, and  the size of the \ci, \twco\ and 
isotope maps is limited to the region mapped in \cii\  by using the \cii\ data cube as a template in GILDAS/CLASS. In this way, we 
produce spectral cubes with the same dimensions and number of pixels.

Then we determine which region of the maps have \textit{\emph{significant}} emission. This can be done naturally by 
using the rms (noise) level of the spectra corresponding to each pixel in the map (i.e., a 3-$\sigma$ level 
detection), or by defining a threshold for the intensities. Since our maps (especially that of the \cii\ line) 
are not homogeneously sampled, the rms level varies among the different pixels. Hence, we prefer to use a 
fraction of the \textit{global} peak (maximum) integrated intensity found among all the channel maps as a 
threshold. This provides a unique value that is used to determine whether the emission of some region (pixel) in 
a particular channel map is significant or not. 

We create binary images for each channel map by assigning a zero to all the pixels with intensity values lower 
than the threshold, and a value of unity to all the pixels that have intensity values larger than or equal to the 
threshold. We use a conservative value of 10\% of the global peak emission for the threshold, which is about ten 
times higher than the noise level in most of the channel maps of all the lines we consider. This conservative 
value is used in order to avoid the association of emission levels in one image that would be considered noise 
in another image.

\begin{figure*}[!ht]

  \hfill\includegraphics[,angle=0,width=0.45\textwidth]{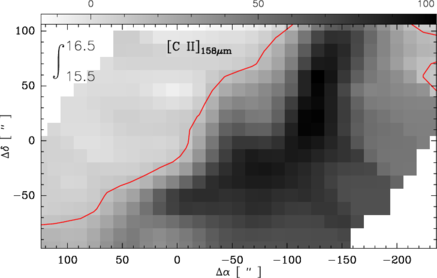}%
  \hfill\includegraphics[angle=0,width=0.45\textwidth]{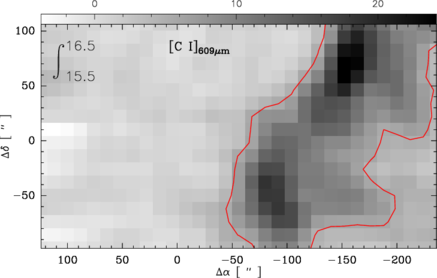}\hspace*{\fill}

  \hfill\includegraphics[,angle=0,width=0.45\textwidth]{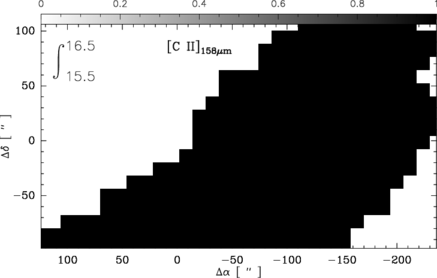}%
  \hfill\includegraphics[,angle=0,width=0.45\textwidth]{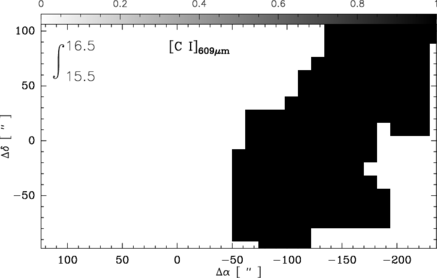}\hspace*{\fill}\\  
  
  \hfill\includegraphics[,angle=0,width=0.45\textwidth]{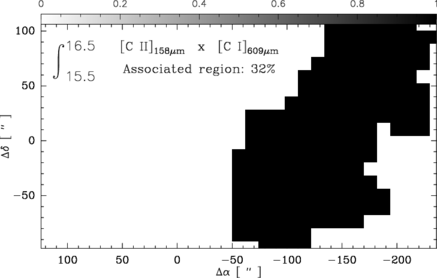}\hspace*{\fill}\\

  \caption{\footnotesize{Example of the steps in the method to estimate the spatial association between two emission lines. The \textit{top} panels shows the channel maps of the 1~\kms\ integrated intensity (in \Kkms) of the \cii~158~\mum\ (\textit{left}) and \ci~609~\mum\ (\textit{right}) lines. The contour lines correspond to the threshold of 10\% of their \textit{global peak intensities} (136.4~\Kkms\ and 54.6~\Kkms\ for \cii\ and \ci, respectively). The \textit{middle} panels are the binary images obtained after applying the intensity threshold to the original channel maps. The \textit{bottom} panel shows the result of multiplying the two binary images, which corresponds to the region where the emission of both \cii\ and \ci\ lines are associated in a particular velocity channel (in this case 15.5--16.5~\kms).}}

  \label{fig:associated-region}
\end{figure*}

Then we multiply each binary channel map image of the two line tracers we want to compare, to see if there are 
regions where the two emissions are associated in that particular velocity channel. The product image would 
contain pixels with 1's in regions where both line tracers have significant emission, and 0's otherwise. Thus, 
adding up all the pixels from the product image, and dividing by the total number of pixels, we obtain the 
fraction of the region mapped where the emission from both lines is associated. An example of this procedure, 
applied to the \cii~158~\mum\ and \ci~609~\mum\ lines, is shown in Fig.~\ref{fig:associated-region}. 

\begin{figure*}[p]

  \hfill\includegraphics[angle=0,width=0.24\textwidth]{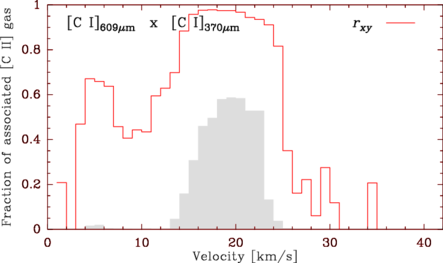}%
  \hfill\includegraphics[angle=0,width=0.24\textwidth]{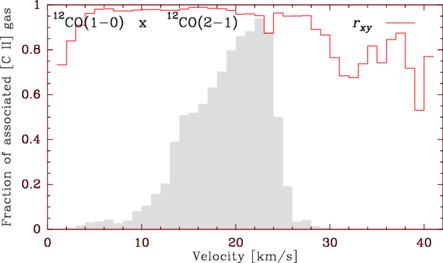}%
  \hfill\includegraphics[angle=0,width=0.24\textwidth]{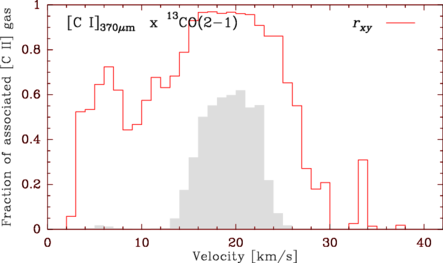}%
  \hfill\includegraphics[angle=0,width=0.24\textwidth]{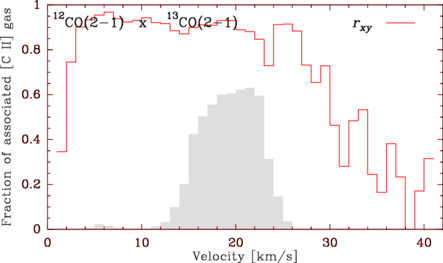} \hspace*{\fill}\\  

  \caption{\footnotesize{Histograms of the fraction of the region mapped where two line tracers are associated at each 1~\kms\ width velocity channel. The corresponding correlation coefficient $R_{xy}$ obtained using Eq.~\ref{eq:corr} for each channel map is shown with a continuous line.}}

  \label{fig:associated-corr-hist}
\end{figure*}

With this method we can estimate the fraction, and where in the region mapped, two line tracers are associated at 
each velocity channel. The fraction of the region mapped can be compared with the correlation coefficient 
described in Appendix~\ref{sec:appendix-B}. Test cases of this are shown in Fig.~\ref{fig:associated-corr-hist}.

\section{\cii\ emission associated with the three gas phases}\label{sec:appendix-D}

\begin{figure*}[!ht]

  \hfill\includegraphics[,angle=0,width=0.48\textwidth]{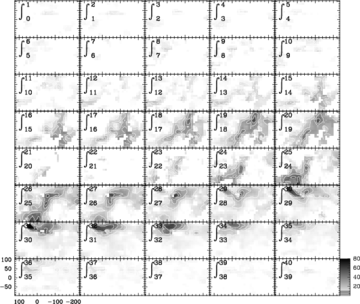}%
  \hfill\includegraphics[angle=0,width=0.48\textwidth]{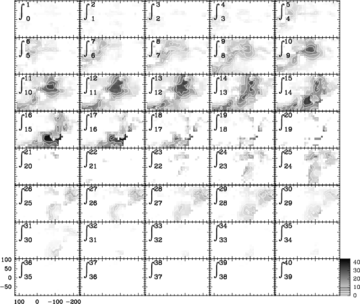}\hspace*{\fill}

  \hfill\includegraphics[,angle=0,width=0.48\textwidth]{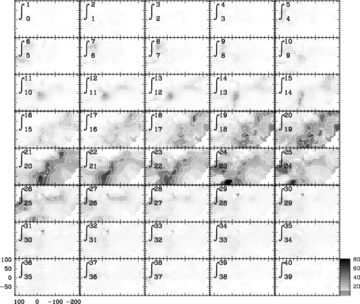}\hspace*{\fill}\\

  \caption{\footnotesize{Velocity channel maps at 1~\kms\ width of the \cii\ emission (in \Kkms), 
  associated with the \hii\ (\textit{top left}), \h1\ (\textit{top right}), and \hh\ (\textit{bottom}) gas phase, 
  as estimated with model (2) from Table~\ref{tab:CII-phases}. Contours are  20\%, 40\%, 60\%, 80\%, 
  and 100\% of the respective peak emissions}}

  \label{fig:CII-H2mass-phases}
\end{figure*}

Subtracting the scaled up spectra of \ci~609~\mum, \twco~$J=1\to0$, and the velocity-resolved optical depth $\tau$(\h1) from the \cii\ spectra, we obtained a residual \cii\ emission that should be mostly associated 
with the ionized hydrogen gas (\hii). When subtracting the maximum (channel by channel) between the tracers of 
the \hh\ gas (e.g., \ci~609~\mum\ and \twco~$J=1\to0$, or \ci~609~\mum\ and \ceio~$J=2\to1$) from the original 
\cii\ spectra, we obtain a second residual \cii\ emission that is expected to be mostly associated with the 
neutral (\h1) and ionized atomic gas (\hii). The difference between these residual spectra and the original \cii\ 
spectra gives the \cii\ emission that is mostly associated with the \hh\ gas. Subtracting the the \cii\ emission that is mostly associated with the \hh\ gas and the 
first residual spectra associated 
with \hii\  from the original \cii\ spectra would lead to a third residual \cii\ emission that is mostly 
associated with the neutral atomic hydrogen gas, \h1. 
The corresponding column densities can be estimated assuming the LTE conditions and Eq. 
\ref{eq:CII-column-LTE}. The velocity channel maps of the residual \cii\ emission associated to each of the gas phases, as estimated with model (2) from Table~\ref{tab:CII-phases}, is shown in Fig.~\ref{fig:CII-H2mass-phases}.

\end{appendix}

\end{document}